\RequirePackage{ifpdf}
\ifpdf % We are running pdfTeX in pdf mode
\documentclass[pdftex]{sigma}
\else
\documentclass{sigma}
\fi

\numberwithin{equation}{section}

\numberwithin{proposition}{section}
\numberwithin{lemma}{section}
\numberwithin{remark}{section}

{\theoremstyle{definition}
\newtheorem{notation}{Notation}
}

\numberwithin{notation}{section}

\def\func#1{\mathop{\rm #1}\nolimits}%

\begin{document}

\allowdisplaybreaks

\renewcommand{\thefootnote}{$\star$}

\renewcommand{\PaperNumber}{082}

\FirstPageHeading

\ShortArticleName{Discrete-Time Goldf\/ishing}

\ArticleName{Discrete-Time Goldf\/ishing\footnote{This
paper is a contribution to the Proceedings of the Conference ``Symmetries and Integrability of Dif\/ference Equations (SIDE-9)'' (June 14--18, 2010, Varna, Bulgaria). The full collection is available at \href{http://www.emis.de/journals/SIGMA/SIDE-9.html}{http://www.emis.de/journals/SIGMA/SIDE-9.html}}}

\Author{Francesco CALOGERO}

\AuthorNameForHeading{F.~Calogero}

\Address{Physics Department, University of Rome ``La Sapienza'',\\
Istituto Nazionale di Fisica Nucleare, Sezione di Roma, Italy}
\Email{\href{mailto:francesco.calogero@roma1.infn.it}{francesco.calogero@roma1.infn.it}, \href{mailto:francesco.calogero@uniroma1.it}{francesco.calogero@uniroma1.it}}

\ArticleDates{Received May 04, 2011, in f\/inal form July 29, 2011;  Published online August 23, 2011}

\Abstract{The original \textit{continuous-time} ``goldf\/ish'' dynamical system is
characterized by  two neat formulas,
the f\/irst of which provides the $N$ Newtonian equations of motion of this
dynamical system, while the second provides the solution of the
corresponding initial-value problem.
Several other, more general, \textit{solvable} dynamical systems ``of goldf\/ish type'' have been identif\/ied over
time, featuring, in the right-hand (``forces'') side of their Newtonian
equations of motion, in addition to other contributions, a
velocity-dependent term such as that appearing in the right-hand side of the
f\/irst formula mentioned above. The \textit{solvable} character of these models
allows detailed analyses of their behavior, which in some cases is quite
remarkable (for instance \textit{isochronous} or \textit{asymptotically
isochronous}). In this paper we introduce and discuss various \textit{discrete-time} dynamical systems, which are as well \textit{solvable}, which
also display interesting behaviors (including \textit{isochrony} and \textit{asymptotic isochrony}) and which reduce to dynamical systems of goldf\/ish
type in the limit when the \textit{discrete-time} independent variable $\ell
=0,1,2,\dots$ becomes the standard \textit{continuous-time} independent
variable $t$, $0\leq t<\infty $.}


\Classification{37J35;  37C27; 70F10; 70H06}

\renewcommand{\thefootnote}{\arabic{footnote}}
\setcounter{footnote}{0}

\section{Introduction}\label{section1}

The original ``goldf\/ish'' dynamical system \cite{C1978,C2001a} is
characterized by the system of $N$ Newtonian equations of motion
\begin{subequations}
\label{Golda}
\begin{gather}
\ddot{z}_{n}=\sum_{k=1,k\neq n}^{N}\frac{2 \dot{z}_{n} \dot{z}_{k}}{z_{n}-z_{k}},\qquad n=1,\dots,N,  \label{NewtGolda}
\end{gather}
and by the following neat prescription yielding the solution of the
corresponding initial-value problem: the $N$ values of the dependent
variables $z_{n}\equiv z_{n}( t) $ at time $t$ are the $N$
solutions of the algebraic equation (for the unknown $z$)
\begin{gather}
\sum_{k=1}^{N}\frac{\dot{z}_{k}( 0) }{z-z_{k}( 0) }=\frac{1}{t},  \label{SolvGolda}
\end{gather}
\end{subequations}
i.e.\ the $N$ roots of the polynomial equation of degree $N$ in the variable
$z$ that obtains by multiplying this formula by the polynomial $
\prod\limits_{j=1}^{N}[ z-z_{j}( 0) ] $.

%\textit{Notation 1.1}.
\begin{notation}\label{notation1.1}
Here and hereafter $N$ is an arbitrary \textit{positive integer} (generally $N\geq 2$), superimposed dots denote
dif\/ferentiations with respect to the independent variable $t$ (``conti\-nuous
time''), and the~$N$ dependent variables $z_{n}\equiv z_{n}( t) $
may be interpreted as the coordinates of~$N$ point-like unit-mass
particles~-- hence $\dot{z}_{n}$ denote their velocities and $\ddot{z}_{n}$
their accelerations, consistently with the interpretation of~(\ref{NewtGolda}) as a set of Newtonian equations of motion with velocity-dependent forces.
The indices~-- such as $n$, $m$, $j$, $k$~-- generally run from~$1$ to~$N$ (below,
as a~convenient reminder, we often indicate this explicitly; as well as the
exceptions to this rule).
\end{notation}

Hereafter we denote as ``dynamical system of goldf\/ish type'' any dynamical
system charac\-te\-ri\-zed by Newtonian equations of motion featuring in their
right-hand sides~-- which have, in the Newtonian context, the signif\/icance of
``forces''~-- a~velocity-dependent term such as that appearing in the
right-hand side of~(\ref{NewtGolda}) (of course in addition to other terms).
Let us also emphasize that the dynamical system~(\ref{Golda}) is the
simplest model belonging to the Ruijsenaars--Schneider \textit{integrable}
class~\cite{RS1986,C2001}.

For instance a simple extension of the above model (reducing to it for $\omega =0$) is characterized by the Newtonian equations of motion:
\begin{subequations}
\label{Goldb}
\begin{gather}
 \ddot{z}_{n}=( 1-2 \alpha ) i \omega \dot{z}_{n}+\alpha
( \alpha -1)  \omega ^{2} z_{n}  \notag \\
\phantom{\ddot{z}_{n}=}{} +\sum_{m=1,m\neq n}^{N}\frac{2 ( \dot{z}_{n}+i \alpha \omega
z_{n}) ( \dot{z}_{m}+i \alpha \omega z_{m}) }{z_{n}-z_{m}},\qquad n=1,\dots,N.  \label{NewtGoldb}
\end{gather}
The corresponding solution of the initial-value problem is again given by a
simple rule: the $N$ values of the dependent variables $z_{n}\equiv
z_{n}( t) $ at time $t$ are related by the formula%
\begin{gather}
z_{n}( t) =\zeta _{n}( t) \exp ( -i \alpha
 \omega t) ,\qquad n=1,\dots,N,  \label{SolvGold1a}
\end{gather}
to the $N$ solutions $\zeta _{n}( t) $ of the algebraic equation
(for the unknown $\zeta $)
\begin{gather}
\sum_{k=1}^{N}\frac{\dot{z}_{k}( 0) +i\alpha \omega
z_{k}( 0) }{\zeta -z_{k}( 0) }=\frac{i\omega}{\exp
(i\omega t) -1}  \label{SolvGold1b}
\end{gather}%
(which again gets transformed into a polynomial equation of degree $N$ in $\zeta $ after multiplication by the polynomial $\prod\limits_{j=1}^{N}%
[ \zeta -z_{j}( 0) ] $); or, equivalently but more
directly, the $N$ values of the dependent variables $z_{n}\equiv z_{n}(
t) $ at time $t$ are the $N$ solutions $z_{n}( t) $ of the
algebraic equation (for the unknown~$z$)
\begin{gather}
\sum_{k=1}^{N}\frac{\dot{z}_{k}( 0) +i\alpha \omega
z_{k}( 0) }{z-z_{k}( 0) \exp ( -i\alpha
\omega t) }=\frac{i\omega \exp ( i\alpha \omega t)}{\exp ( i\omega t) -1}.  \label{Solz1}
\end{gather}
\end{subequations}

%\textit{Notation 1.2}.
\begin{notation}\label{notation1.2}
 Here and hereafter $i$ is the imaginary unit, $i^{2}=-1$, $\omega $ is an arbitrary constant (dimensionally, an inverse
time), and $\alpha $ is an arbitrary (dimensionless) constant.
Clearly~-- unless both $\omega $ and $\alpha \omega $ are both imaginary, $\func{Re}
( \omega ) =\func{Re} ( \alpha \omega ) =0$~-- the
time-evolution of this system takes place in the \textit{complex} $z$-plane,
i.e.\ the dependent variables $z_{n}\equiv z_{n}( t) $ are
\textit{complex}; but it may as well be viewed as describing the evolution
of $N$ point-like particles moving in the \textit{real} $xy$-plane~-- whose
positions at time $t$ are characterized by the (\textit{real}) Cartesian
coordinates $x_{n}\equiv x_{n}( t) ,$ $y_{n}\equiv y_{n}(
t) $~-- by setting $z_{n}( t) =x_{n}( t)
+iy_{n}( t) $; and one of the remarkable features of the
resulting \textit{real} model is the possibility to write its Newtonian
equations of motion in \textit{covariant}~-- i.e., \textit{rotation-invariant}~-- form, see Chapter 4 of~\cite{C2001}. Hereafter we
generally refer to this model, and its generalizations, see below, in their
\textit{complex} versions.
\end{notation}

%\textit{Remark 1.1}.
\begin{remark}\label{remark1.1}
Clearly for $\omega =0$ this model, (\ref{Goldb}),
reduces to the previous model (\ref{Golda}). For $\omega \neq 0$ the time
evolution of this model, (\ref{Goldb}), depends mainly on the values of the
two constants~$\omega $ and~$\alpha \omega $, as displayed by its solution,
see (\ref{Solz1}). If both these constants are \textit{real}, $\func{Im}( \omega ) =\func{Im}( \alpha \omega ) =0$ (hence as
well $\func{Im}( \alpha ) =0$), the time evolution of this model
is \textit{confined}, indeed \textit{completely periodic} if the \textit{real} number $\alpha $ is \textit{rational}, while if $\alpha $ is \textit{irrational} it is \textit{multiply periodic}, being a nonlinear
superposition of two periodic evolutions with the two noncongruent periods
$T=2\pi /\left\vert \omega \right\vert $ and~$T/\alpha$. Note that these
outcomes obtain for \textit{generic} initial data: hence, for~$\alpha $
\textit{rational}, $\alpha =q/p$ with~$q$ and~$p$ \textit{coprime integers}
(and $p>0$), this system is \textit{isochronous}, its \textit{generic}
solutions being \textit{completely periodic} with period~$pT$~-- or possibly
with a period which is a, generally small, \textit{integer multiple} of $pT$: indeed, when the equation~(\ref{Solz1}) is itself periodic with period $pT$, the \textit{unordered} \textit{set} of its $N$ roots is clearly periodic
with the same period $pT$, but the periodicity of the time-evolution of each
individual coordinate $z_{n}( t) $ may then be a, generally
small, \textit{integer multiple} of~$pT$ due to the possibility that
dif\/ferent roots get exchanged through the time evolution (for a discussion
of this phenomenology~-- including a justif\/ication of the assertion that the
relevant integer multiple of~$pT$ is generally small~-- see~\cite{GS2005}).

On the other hand, if $\omega $ is \textit{real} but $\alpha $ is imaginary,
say $\alpha \omega =i\gamma $ with $\gamma $ \textit{real} and \textit{nonvanishing}, then clearly in the remote future~-- i.e., as $t\rightarrow
\infty $, and up to relative corrections of order $\exp ( -\vert
\gamma \vert t) $~-- all the $N$ coordinates $z_{n}(
t) $ tend to the origin, $z_{n}( \infty ) =0,$ if $\gamma
<0 $, while if $\gamma >0$ they all diverge (see~(\ref{SolvGold1a}) and~(\ref{SolvGold1b})).

If instead $\omega $ is \textit{not real}, $\func{Im}( \omega )
\neq 0$, then in the remote future (i.e., as $t\rightarrow \infty $, and up
to relative corrections of order $\exp ( -\vert \func{Im}(
\omega ) \vert t) $) the $N$ solutions of (\ref{SolvGold1b}) become asymptotically, if $\func{Im}( \omega ) >0$, the $N$
solutions $\zeta _{n}=\zeta _{n}( \infty ) $ of the \textit{time-independent} polynomial equation of order~$N$ in~$\zeta $
\begin{gather*}
\sum_{k=1}^{N}\frac{\dot{z}_{k}( 0) +i\alpha \omega
z_{k}( 0) }{\zeta -z_{k}( 0) }=-i\omega,
\end{gather*}
while if instead $\func{Im}( \omega ) <0$ the equation (\ref{SolvGold1b}) becomes, in the remote future, the \textit{time-independent}
polynomial equation
%\begin{subequations}
\begin{gather*}
\sum_{k=1}^{N}\frac{\dot{z}_{k}( 0) +i \alpha \omega
z_{k}( 0) }{\zeta -z_{k}( 0) }=0,
\end{gather*}
hence $N-1$ of the solutions of (\ref{SolvGold1b}) tend asymptotically to
the $N-1$ solutions of this equation (polynomial of degree $N-1$ in $\zeta $) and one of them approaches asymptotically the \textit{diverging} coordinate%
\begin{gather*}
\zeta _{\rm asy}( t) =\exp ( i \omega t) \sum_{k=1}^{N}
[ \dot{z}_{k}( 0) +i \alpha \omega z_{k}( 0) ].
\end{gather*}
%\end{subequations}
Note that this implies (see (\ref{SolvGold1a})) that, if $\func{Im}(
\omega ) >0$ but $\alpha \omega $ is \textit{real}, $\alpha \omega
=\rho $ with $\rho $ \textit{real} and \textit{nonvanishing}, then the model
(\ref{Goldb}) is \textit{asymptotically isochronous}, its \textit{generic}
solutions beco\-ming, in the remote future, \textit{completely periodic} with
period $2\pi /\vert \rho \vert $, up to corrections vanishing
exponentially as $t\rightarrow \infty $ (for a more detailed discussion of
the notion of \textit{asymptotic isochrony} see Chapter 6, entitled
``Asymptotically isochronous systems'', of~\cite{C2008}).
\end{remark}

For $\omega =0$ (i.e., when the model~(\ref{Goldb}) reduces to (\ref{Golda})) it is possible to restrict consideration to \textit{real} dependent
variables~$z_{n}$, but even then it is more interesting \textit{not} to do
so, so that the time evolution takes place in the plane rather than on the
real line: see the remarkable behavior of this dynamical system in this case
(``the game of musical chairs''), as detailed in Section~4.2.4 of~\cite{C2001}.
Hence let us reiterate that we always consider the dependent variables $z_{n}$ to be \textit{complex} numbers, both in the \textit{continuous-time}
case, $z_{n}\equiv z_{n}( t) $, $0\leq t<\infty $, and (see
below) in the \textit{discrete-time} case, $z_{n}\equiv z_{n}(\ell )$, $\ell =0,1,2,\dots$.

Another large class of \textit{solvable} dynamical systems ``of goldf\/ish
type'' is characterized by the equations of motion
\begin{gather}
\ddot{z}_{n}=a_{1}\dot{z}_{n}+a_{2}+a_{3}z_{n}-2(N-1)a_{4}z_{n}^{2}+
\sum_{m=1,\;m\neq n}^{N}( z_{n}-z_{m}) ^{-1}\big[ 2\dot{z}_{n}
\dot{z}_{m}   \notag \\
  \phantom{\ddot{z}_{n}=}{} + ( a_{5}+a_{6}z_{n} )  ( \dot{z}_{n}+\dot{z}
_{m} ) +a_{7}z_{n} ( \dot{z}_{n}z_{m}+\dot{z}_{m}z_{n} )
+2\big( a_{8}+a_{9}z_{n}+a_{10}z_{n}^{2}+a_{4}z_{n}^{3}\big) \big],
\notag \\
\phantom{\ddot{z}_{n}=}{} n=1,\dots,N,  \label{GenGoldfish}
\end{gather}
featuring 10 arbitrary constants (see   \cite[equation~(2.3.3-2)]{C2001}).
In this case the \textit{solvability} is achieved by identifying the $N$
dependent variables $z_{n} ( t ) $ with the $N$ roots of a
time-dependent polynomial $\psi  ( z,t ) $ of degree~$N$ in~$z$
satisfying a \textit{linear} second-order PDE in the two independent
variables~$z$ and~$t$.

For an explanation of the origin of the name ``goldf\/ish'' attributed to these
models see Section~1.N of~\cite{C2008} and the literature cited there. In
this book~\cite{C2008} (see in particular its Section~4.2.2, entitled
``Goldf\/ishing'', and the papers referred to there) several other \textit{solvable} models ``of goldf\/ish type'' are reported, including \textit{isochronous} ones (i.e., models featuring solutions which are \textit{completely periodic with a period independent of the initial data}). A~few additional models of goldf\/ish type have been identif\/ied more recently \cite{C2011d,C2011,C2011c}.

The most remarkable aspect of these dynamical systems is their \textit{solvability}, namely the possibility to solve their initial-value problems
by \textit{algebraic} operations, amounting generally to f\/inding the $N$
eigenvalues of an $N\times N$ explicitly known time-dependent matrix (see
below), or equivalently to f\/inding the $N$ roots of an explicitly known
time-dependent polynomial of degree~$N$ (see for instance~(\ref{SolvGolda})
and (\ref{Solz1})). Quite interesting is also the identif\/ication of \textit{multiply periodic}, \textit{completely periodic}, or even \textit{isochronous} or \textit{asymptotically isochronous} cases.

In the present paper we present various \textit{discrete-time} dynamical
systems ``of goldf\/ish type'', so denoted because all these models reduce, in
the limit when the \textit{discrete-time} independent variable $\ell
=0,1,2,\dots$ becomes \textit{continuous}, to \textit{continuous-time}
dynamical systems of goldf\/ish type. All these models are moreover \textit{solvable}, i.e.\ the solution of their initial value problems can be
achieved by f\/inding the $N$ eigenvalues $z_{n} ( \ell  ) $ of $N\times N$ matrices explicitly known in terms of the initial data and of the
\textit{discrete-time} independent variable~$\ell $; or equivalently by
f\/inding the $N$ roots $z_{n} ( \ell  ) $ of a polynomial, of degree
$N$ in the complex variable~$z$, as well explicitly known in terms of the
initial data and of the \textit{discrete-time} independent variable~$\ell$.
Some of these models feature interesting behaviors, even \textit{isochrony}
or \textit{asymptotic isochrony}. Two of these models (see Subsection~\ref{section2.1}
and~\ref{section2.2}) were treated in the paper~\cite{C2011a}, which has not been
published because~-- after it was submitted for publication but before
getting any feedback~-- new solvable models were identif\/ied and it was
therefore considered preferable to report all these models in a single
paper, this one. The main properties of each of these \textit{discrete-time}
models are reported in Section~\ref{section2}, and proven in Section~\ref{section3}. These properties
include the display of the equations of motion of these \textit{discrete-time} models, the solution of their initial-value problems, a terse discussion
(for the f\/irst three models) of their behavior including the possibility
that for special values of some of their parameters they possesses \textit{periodic} or \textit{multiperiodic} solutions or even display \textit{%
isochrony} or \textit{asymptotic isochrony}, and some mention of their
\textit{continuous-time} limits. Section~\ref{section4} entitled ``Outlook'' concludes
the paper: in it a general framework is outlined which might allow the
identif\/ication of additional solvable \textit{discrete-time} models. And
some mathematical developments are conf\/ined to two appendices.

These f\/indings are congruent with the recent surge of interest for \textit{discrete-time} evolutions~-- in particular, such evolutions which are in some
sense \textit{integrable} or even \textit{solvable}. Given the large body of
research devoted to these topics over the last two decades, our reference to
the relevant literature shall be limited to citing the following surveys:
\cite{V1991,CN1999,S2003,R2006,BS2008}. But a~special mention must be made of the seminal papers by Nijhof\/f, Ragnisco,
Kuznetsov and Pang, see~\cite{NRK1996} as well as the earlier paper \cite{NP1994}, where \textit{discrete-time} versions were introduced of the
well-known integrable ``Ruijsenaars--Schneider'' and ``Calogero--Moser'' dynamical
systems; as well as of the paper by Suris~\cite{S2005}, which treats
specif\/ically a \textit{discrete-time} version of the original goldf\/ish
model. Some results of these papers refer to models whose equations of
motion feature trigonometric/hyperbolic or even elliptic functions, and are
therefore more general than those treated in the present paper, whose
equations of motion only feature rational functions (see below); on the
other hand the f\/indings reported below include more general models than
those previously treated, demonstrate the \textit{solvability }of these
models by an approach somewhat dif\/ferent from those previously employed,
and, most signif\/icantly, display the possible emergence of remarkable
phenomenologies~-- including \textit{periodicity} and even \textit{isochrony}
or \textit{asymptotic isochrony}, see below~-- not previously
identif\/ied for this kind of \textit{discrete-time} dynamical systems.

Let us also mention that the approach developed below also allows to
identify and investigate \textit{discrete-time} variants of another class of
solvable \textit{continuous-time} dynamical systems, the prototype of which
is characterized by the Newtonian equations of motion
\begin{gather*}
\ddot{z}_{n}=\sum_{k=1,\;k\neq n}^{N}\frac{c}{ ( z_{n}-z_{k} ) ^{3}},\qquad n=1,\dots,N
\end{gather*}
(with $c$ an arbitrary constant), instead of~(\ref{NewtGolda}). But in this
paper we merely indicate, at the appropriate point, how to proceed in this
direction, postponing a complete treatment of this development to a separate
paper.

And let us f\/inally pay tribute to Olshanetsky and Perelomov who were the
f\/irst to show, more than~35 years ago, that the time-evolution of a
nontrivial many-body system could be usefully identif\/ied with the evolution
of the eigenvalues of a matrix itself evolving in a much simpler, explicitly
solvable, manner: see \cite{OP1976} and their other papers referred to
in Section~2.1.3.2 of~\cite{C2001}, entitled ``The technique of solution of
Olshanetsky and Perelomov''. The present paper extends their approach to the
\textit{discrete-time} context.

\section{Results}\label{section2}

In this section we report the main results of this paper; they are then
proven in Section~\ref{section3}.

\begin{notation}\label{notation2.1}
%\textit{Notation 2.1}.
Hereafter the dependent variables are indicated again
as $z_{n}$, but they are now functions, $z_{n}\equiv z_{n} ( \ell
 ) $, of the \textit{discrete-time} variable $\ell $ taking the \textit{integer} values $\ell =0,1,2,\dots$; and superimposed tildes indicate
generally a \textit{unit} increase of the independent variable $\ell $, for
instance $\tilde{z}_{n}\equiv z_{n} ( \ell +1 )$, $\widetilde{\tilde{z}}_{n}\equiv z_{n} ( \ell +2 ) $. Hereafter $\delta _{nm}$ is the
standard Kronecker symbol, $\delta _{nm}=1$ if $n=m$, $\delta _{nm}=0$ if $n\neq m$, and underlined quantities are $N$-vectors, for instance $\underline{z}\equiv ( z_{1},\dots,z_{N} ) $. For the remaining
notation we refer to Notation~\ref{notation1.1}, see above, and to specif\/ic
indications given case-by-case below.
\end{notation}

As reported in this section and explained in Sections~\ref{section3} and~\ref{section4}, the solvable
models considered in this paper generally feature three \textit{equivalent}
versions of the second-order ``equations of motion'' characterizing their
evolution in \textit{discrete-time}. The treatment of the f\/irst model given
in Subsections~\ref{section2.1} and~\ref{section3.1} is somewhat more detailed than that provided for
the other models in the subsequent subsections where, to avoid repetitions,
we often refer to the treatments provided in Subsections~\ref{section2.1} and~\ref{section3.1}. And
already in this section, as well as in Section~\ref{section3}, we often
take advantage~-- to simplify the presentation of some results~-- of \textit{identities} and \textit{lemmata} collected in Appendix~\ref{appendixA}.

\subsection{First model}\label{section2.1}

The f\/irst model is def\/ined by the following second-order \textit{discrete-time} equations of motion: the $N$ values of the twice-updated
variables $\widetilde{\tilde{z}}_{n}\equiv z_{n} ( \ell +2 ) $ are
given, in terms of the $2N$ values of the variables $z_{m}\equiv z_{m} (
\ell  )$, $\tilde{z}_{m}\equiv z_{m} ( \ell +1 ) $, by the $N$
roots of the following (\textit{single}) algebraic equation in the unknown~$z $,
\begin{subequations}
\label{DiscreteGold}
\begin{gather}
\sum_{k=1}^{N}\left[ \left( \frac{\tilde{z}_{k}-az_{k}}{z-a\tilde{z}_{k}}
\right) \prod\limits_{j=1,\; j\neq k}^{N}\left( \frac{\tilde{z}_{k}-az_{j}}{
\tilde{z}_{k}-\tilde{z}_{j}}\right) \right] =1,  \label{DiscreteGolda}
\end{gather}
which clearly amounts to a polynomial equation of degree~$N$ in this
variable $z$ (as it is imme\-diate\-ly seen by multiplying this equation by the
polynomial $\prod\limits_{m=1}^{N} ( z-a\tilde{z}_{m} ) $). Here
and below $a$ is an arbitrary (dimensionless, nonvanishing) constant. A~neater version of this formula is easily obtained by multiplying it by~$a$
and by then using the identity (\ref{Identity3b}) with $\eta _{n}=a\tilde{z}_{n}$,
$\zeta _{n}=a^{2}z_{n}$, $n=1,\dots,N$. It reads
\begin{gather}
\prod\limits_{j=1}^{N}\left( \frac{z-a^{2}z_{j}}{z-a\tilde{z}_{j}}\right)
=1+a.  \label{DiscreteGoldb}
\end{gather}
\end{subequations}

An \textit{equivalent} formulation of this model is provided by the
following system of $N$ polynomial equations of degree~$N$ for the
twice-updated coordinates $\widetilde{\tilde{z}}_{n}\equiv z_{n} ( \ell
+2 ) $:
\begin{subequations}
\label{Equiv}
\begin{gather}
\sum_{k=1}^{N}\left[ \left( \frac{\widetilde{\tilde{z}}_{k}-a \tilde{z}_{k}}{\tilde{z}_{k}-a
z_{n}}\right)  \prod\limits_{j=1,\;j\neq k}^{N}\left( \frac{\widetilde{\tilde{z}}_{j}-a\tilde{z}_{k}}{\tilde{z}_{j}-\tilde{z}_{k}}
\right) \right] =a^{N-1},\qquad n=1,\dots,N.  \label{Equiva}
\end{gather}
Again, a neater version of this formula is easily obtained by dividing it by
$a$ and by then using the identity (\ref{Identity3b}), now with $z=a^{2}z_{n}$, $\eta _{n}=a~\tilde{z}_{n}$, $\zeta _{n}=\widetilde{\tilde{z}}_{n}$, $n=1,\dots,N$. It reads%
\begin{gather}
\prod\limits_{j=1}^{N}\left( \frac{\widetilde{\tilde{z}}_{j}-a^{2} z_{n}}{\tilde{z}_{j}-a z_{n}}\right)
= ( 1+a ) a^{N-1},\qquad n=1,\dots,N.
\label{Equivb}
\end{gather}
\end{subequations}

And a third, \textit{equivalent} version of this model is provided by the
following system of $N$ polynomial equations of degree $N$ for the
twice-updated coordinates $\widetilde{\tilde{z}}_{n}\equiv z_{n} ( \ell
+2 ) $:
\begin{gather}
\prod\limits_{j=1}^{N}\left( \frac{\widetilde{\tilde{z}}_{j}-a \tilde{z}_{n}
}{az_{j}-\tilde{z}_{n}}\right) =-a^{N-1},\qquad n=1,\dots,N.
\label{EqMotMod1ter}
\end{gather}

The similarities and dif\/ferences among these three sets of ``equations of
motion'', (\ref{DiscreteGold}), (\ref{Equiv}) and (\ref{EqMotMod1ter}), are
remarkable: let us reemphasize that they in fact yield the \textit{same}
evolution in discrete-time of the $N$ coordinates $z_{n}\equiv z_{n} (
\ell  ) $. Particularly remarkable is their similarity in the special $a=1$ case, when the 3 versions~(\ref{DiscreteGoldb}), (\ref{Equivb}) and~(\ref{EqMotMod1ter}) of the equations of motion read as follows:
%\begin{subequations}
\begin{gather*}
\prod\limits_{j=1}^{N}\left( \frac{\widetilde{\tilde{z}}_{n}-z_{j}}{\widetilde{\tilde{z}}_{n}-\tilde{z}_{j}}\right) =2, \qquad n=1,\dots,N,
\\
\prod\limits_{j=1}^{N}\left( \frac{\widetilde{\tilde{z}}_{j}-z_{n}}{\tilde{z}_{j}-z_{n}}\right) =2,\qquad n=1,\dots,N,
\\
\prod\limits_{j=1,\; j\neq n}^{N}\left( \frac{\widetilde{\tilde{z}}_{j}-\tilde{
z}_{n}}{z_{j}-\tilde{z}_{n}}\right) =-1,\qquad n=1,\dots,N.
\end{gather*}
%\end{subequations}
The last of these three systems coincides with equation~(1.8) of \cite{S2005}.

%\textit{Remark 2.1.1}.
\begin{remark}\label{remark2.1.1}
This model, see (\ref{DiscreteGold}), (\ref{Equiv})
and (\ref{EqMotMod1ter})~-- as the original goldf\/ish model, see (\ref{Goldb})~-- is \textit{invariant} under an \textit{arbitrary} rescaling of the
dependent variables, $z_{n}\Rightarrow c z_{n}$ with $c$ an \textit{arbitrary constant};
including the special case $c=\exp  ( i \gamma  ) $ with $\gamma $ an arbitrary \textit{real} constant, corresponding
to an overall \textit{rotation} around the origin in the \textit{complex} $z$-plane.
\end{remark}

The solution of the initial-value problem for this model is given by the
following

%\textit{Proposition 2.1.1}.
\begin{proposition}\label{proposition2.1.1}
The $N$ values $z_{n} ( \ell  ) $ of
the dependent variables at the discrete time $\ell $ are the $N$ eigenvalues of the $N\times N$ matrix
\begin{subequations}
\label{U}
\begin{gather}
U_{nm} ( \ell  ) =\delta _{nm} z_{n} ( 0 ) a^{\ell
}+v_{m} ( 0 ) \frac{a^{\ell }-1}{a-1},\qquad n,m=1,\dots,N
\label{UnmMod1}
\end{gather}
with
\begin{gather}
v_{m}\equiv v_{m} ( \underline{z},\underline{\tilde{z}} ) =\frac{
\prod\limits_{j=1}^{N} ( \tilde{z}_{j}-a z_{m} ) }{\prod\limits_{j=1,\; j\neq m}^{N} [ a  ( z_{j}-z_{m} )  ] }
,\qquad m=1,\dots,N,  \label{y}
\end{gather}
\end{subequations}
where of course $v_{m} ( 0 ) $ indicates the value of $v_{m} (
\underline{z},\underline{\tilde{z}} ) $ corresponding to the initial
data $\underline{z}=\underline{z} ( 0 ) ,$ $\underline{\tilde{z}}=
\underline{\tilde{z}} ( 0 ) \equiv \underline{z} ( 1 ) $.

A neater, \textit{equivalent} formulation of this finding~-- obtained from \eqref{U} via Lemma~{\rm \ref{lemmaA4}} with $\zeta _{n}=z_{n} ( 0 ) a^{\ell
}$ and $\eta _{m}=v_{m} ( 0 )  ( a^{\ell }-1 ) / (
a-1 ) $~-- states that the $N$ coordinates $z_{n} ( \ell  ) $
are the $N$ solutions of the following algebraic equation in $z$:
\begin{gather*}
\sum_{k=1}^{N}\left\{ \left[ \frac{z_{k} ( 1 ) -a z_{k} (
0 ) }{z-a^{\ell } z_{k} ( 0 ) }\right] \prod\limits_{j=1,\;j\neq k}^{N}\left[ \frac{z_{j} ( 1 ) -az_{k} ( 0 ) }{az_{j} ( 0 ) -a z_{k}(0) }\right] \right\} =\frac{a-1}{a^{\ell }-1}.
\end{gather*}
And another, even neater, \textit{equivalent} formulation~-- obtained from
this via the identity~\eqref{Identity3b} with $z$ replaced by $za^{1-\ell }$,
$\eta _{k}=az_{k}(0) $, $\zeta _{j}=\tilde{z}_{j}(0)
=z_{j}(1) $~-- states that the coordinates $z_{n} ( \ell) $ are the $N$ solutions of the following algebraic equation in $z$:
\begin{gather}
\prod\limits_{k=1}^{N}\left[ \frac{z-a^{\ell -1} z_{k}(1) }{z-a^{\ell }z_{k}(0) }\right]
=\frac{a^{\ell -1}-1}{a^{\ell }-1}.  \label{SolMod1}
\end{gather}
The last two equations become of course polynomial equations of degree $N$
in $z$ after multiplication by the product $\prod\limits_{j=1}^{N}[z-a^{\ell } z_{j}(0)] $.
\end{proposition}

These formulas are also valid for $a=1$ (by taking the obvious limit, i.e.\
replacing $( a^{p}-1)/( a-1) $ with~$p$). If instead $\vert a\vert <1$, then clearly for all (\textit{positive}) values
of $\ell $ the matrix $U( \ell ) $ is bounded and $U_{nm}(
\infty ) =v_{m}(0) /( 1-a) ;$ hence for all
values of $\ell $ the $N$ coordinates $z_{n}( \ell ) $ are
bounded and $N-1$ of them vanish as $\ell \rightarrow \infty $ while one of
them tends to the value
\begin{gather*}
z_{\rm asy}=( 1-a ) ^{-1}\sum_{k=1}^{N}v_{k}(0) =\frac{1}{1-a} \sum_{k=1}^{N} [ z_{k}(1) -az_{k}(0)
 ] ,
\end{gather*}
see (\ref{y}) and the identity (\ref{Identity4}) (with $\eta
_{k}=a~z_{k}(0) $, $\zeta _{j}=\tilde{z}_{j}(0)
=z_{j}(1) $). If $a\neq 1$ but it has \textit{unit} modulus,
\begin{gather}
a=\exp  ( 2\pi i\lambda  )  \label{alanda}
\end{gather}
with $\lambda $ \textit{real} and \textit{not integer}, then clearly the
matrix $U( \ell ) $ is again, for all values of $\ell$,
bounded; and if moreover $\lambda $ is a (strictly, i.e.\ non integer)
\textit{rational} number,
%\begin{subequations}
\begin{gather}
\lambda =\frac{K}{L},\qquad a=\exp \left( \frac{2\pi iK}{L}\right),
\label{landarat}
\end{gather}
with $K$ and $L$ two \textit{coprime integers} and $L>1$, then clearly the
matrix $U( \ell ) $ is \textit{periodic} with period $L$,
\begin{gather*}
U ( l+L ) =U ( l ) ,
\end{gather*}
%\end{subequations}
hence the (unordered) set of its $N$ eigenvalues $z_{n}( \ell ) $
is as well \textit{periodic} with period $L$. This shows that in this case,
see (\ref{landarat}), the \textit{discrete-time} goldf\/ish model, see (\ref{DiscreteGold}) or~(\ref{Equiv}) or~(\ref{EqMotMod1ter}), is \textit{isochronous}. On the other hand if $\lambda $ is \textit{real} and \textit{irrational}, then clearly the time evolution of this \textit{discrete-time}
dynamical system is \textit{not periodic}: indeed, while the right-hand side
of~(\ref{UnmMod1}) (with~(\ref{alanda}) and $\lambda $ \textit{real} and
\textit{irrational}) is periodic (with \textit{unit} period) as a function
of the \textit{real} variable $\tau =\lambda\ell ,$ clearly it is \textit{%
not} periodic as a function of the variable $\ell $ taking the integer
values $\ell =0,1,2,\dots$. (We made this analysis, for convenience,
referring to the coordinates $z_{n}( \ell ) $ as the eigenvalues
of $U( \ell ) $, see~(\ref{U}); of course an analogous discussion
could be made on the basis of the alternative identif\/ication of the
coordinates $z_{n}( \ell ) $ as the $N$ roots of the polynomial
equation~(\ref{SolMod1})~-- whose similarity with (\ref{Solz1}) is in any
case to be noted, see below.)

To explore the transition from the \textit{discrete-time} independent
variable $\ell $ to the \textit{continuous-time} variable $t$ one makes the
formal replacements
\begin{subequations}
\label{Expepsi}
\begin{gather}
\ell  \ \Longrightarrow \ \frac{t}{\varepsilon },\qquad \ell +1 \ \Longrightarrow \
\frac{t+\varepsilon }{\varepsilon }, \qquad \ell +2\ \Longrightarrow \ \frac{t+2\varepsilon }{\varepsilon },  \label{landaepsi}
\\
a\ \Longrightarrow \ 1-i\omega\varepsilon ,  \label{aepsi}
\end{gather}
and (with a slight abuse of notation)%
\begin{gather}
z_{n}( \ell ) \ \Longrightarrow \ z_{n}(t) ,  \notag \\
\tilde{z}_{n}( \ell ) \equiv z_{n} ( \ell +1 )
\ \Longrightarrow \ z_{n}(t) +\varepsilon \dot{z}_{n}(t)+\frac{\varepsilon ^{2}}{2}\ddot{z}_{n}(t) +O\big( \varepsilon
^{3}\big) ,  \notag \\
\widetilde{\tilde{z}}_{n}( \ell ) \equiv  z_{n} ( \ell
+2 ) \ \Longrightarrow \ z_{n}(t) +2\varepsilon \dot{z}_{n}(t)+2\varepsilon ^{2} \ddot{z}_{n}(t)
+O\big( \varepsilon ^{3}\big),  \label{zepsi}
\end{gather}
\end{subequations}
with $\varepsilon $ inf\/initesimal. It is then a matter of standard, if a bit
cumbersome, algebra, to verify that the insertion of this \textit{ansatz},
see (\ref{aepsi}) and (\ref{zepsi}), in~(\ref{DiscreteGoldb}) or~(\ref{Equivb}) or~(\ref{EqMotMod1ter}) yields a trivial identity to order
$\varepsilon ^{0}=1$, while to order $\varepsilon $ it reproduces~(\ref{NewtGoldb}) with $\alpha =1,$ reading
\begin{subequations}
\label{ContModelalphaEq1}
\begin{gather}
\ddot{z}_{n}=-i \omega  \dot{z}_{n}+\sum_{m=1,\; m\neq n}^{N}\frac{2  (
\dot{z}_{n}+i\omega  z_{n} )   ( \dot{z}_{m}+i\omega
z_{m} ) }{z_{n}-z_{m}},\qquad n=1,\dots,N.  \label{ContGoldAlphaEqOne}
\end{gather}
Likewise, the \textit{discrete-time} solution formula (\ref{SolMod1})
becomes, in the \textit{continuous-time} limit,
\begin{gather}
\sum_{k=1}^{N}\frac{\dot{z}_{k}(0) +i\omega z_{k} (
0) }{z-z_{k}(0) \exp ( -i\omega t) }=\frac{i\omega }{1-\exp ( -i\omega t) }\equiv \frac{i\omega \exp
( i\omega t) }{\exp ( i\omega t) -1},
\label{SolvContAlphaEqOne}
\end{gather}
\end{subequations}
which coincides with (\ref{Solz1}) with $\alpha =1$. A terse outline of the
derivation of these results is provided at the end of Subsection~\ref{section3.1}. To
higher order in $\varepsilon $ one would obtain additional relations
satisf\/ied by the solution $z_{n}(t) $ of this \textit{continuous-time} goldf\/ish model, which might alternatively be obtained by
dif\/ferentiating its equations of motion~(\ref{ContGoldAlphaEqOne}).

%\textit{Remark 2.1.2}.
\begin{remark}\label{remark2.1.2}
Clearly, for $\omega $ real and nonvanishing, this
\textit{continuous-time} model, (\ref{ContGoldAlphaEqOne}), is \textit{isochronous}: see~(\ref{SolvContAlphaEqOne}) and/or Remark~\ref{remark1.1}.
This is consistent with the fact that the limiting replacement~(\ref{aepsi})
can be considered to obtain from (\ref{landarat})~-- entailing \textit{isochrony} of the \textit{discrete-time} model~-- by identifying $\varepsilon
\omega $ with $2\pi K/L$ in the context of the replacement (see~(\ref{Expepsi})) of the \textit{unit} interval in the \textit{discrete-time}
model with the \textit{infinitesimal} time interval $\varepsilon $ to make
the transition to the \textit{continuous-time} case.
\end{remark}

%\textit{Remark 2.1.3}.
\begin{remark}\label{remark2.1.3}
At every step of the \textit{discrete-time} evolution
the $N$ values of the twice-updated variables $\widetilde{\tilde{z}}%
_{n}\equiv z_{n} ( \ell +2 ) $ are given, in terms of the $N$
unupdated variables $z_{m}\equiv z_{m}( \ell ) $ and the $N$
once-updated variables $\tilde{z}_{m}\equiv z_{m} ( \ell +1 ) $, as
the $N$ roots of a polynomial, of degree $N$ in its argument $z$, whose
coef\/f\/icients are explicitly def\/ined in terms of the $2N$ unupdated and
once-updated variables: see (\ref{DiscreteGold}) or (\ref{Equiv}) or (\ref{EqMotMod1ter}) (hereafter~-- within this important~Remark~\ref{remark2.1.3}~-- we generally identify, for simplicity, this model only via the version~(\ref{DiscreteGold}) of its equations of motion). Hence at every step of this
\textit{discrete-time} evolution the \textit{unordered} \textit{set} of~$N$
twice-updated variables~$\widetilde{\tilde{z}}_{n}$ is \textit{uniquely}
determined, but \textit{not} the value of \textit{each} of them. This
implies a \textit{qualitative} dif\/ference among the \textit{continuous-time}
respectively the \textit{discrete-time} evolutions described by the
equations of motions~(\ref{ContGoldAlphaEqOne}) (or, more generally, (\ref{NewtGoldb}) and (\ref{GenGoldfish})) respectively by~(\ref{DiscreteGold}):
in contrast to the \textit{continuous-time} case, the \textit{discrete-time}
evolution~(\ref{DiscreteGold}) is \textit{only} deterministic in terms of
the \textit{unordered set} of $N$ coordinates $z_{m}( \ell ) $,
but not for \textit{each} individual coordinate $z_{n}( \ell ) $.
Indeed the \textit{continuous-time} Newtonian equations of motion, see for
instance (\ref{NewtGoldb}), determine \textit{uniquely} the value of the
acceleration $\ddot{z}_{n}(t) $ of the $n$-th moving point in
terms of the $N$ positions $z_{m}(t) $ and the $N$ speeds $\dot{z}_{m}(t) $ of all moving points; and correspondingly, while the
solution formula~(\ref{Solz1}) determines only the \textit{unordered set} of
$N$ values $z_{n}(t) $ as the $N$ roots of a polynomial of
degree~$N$, the value of \textit{each} individual coordinate $z_{n}(t) $ gets then \textit{uniquely} determined by \textit{continuity} in
the time variable $t$. This latter mechanism to identify \textit{uniquely}
the value of the coordinate of \textit{each} moving point is instead missing
for the \textit{discrete-time} evolution (\ref{DiscreteGold}). On the other
hand it is clear that there are appropriate ranges of values of the
parameter $a$ and of the $2N$ initial data~$z_{m}(0)$,~$\tilde{z}%
_{m}(0) \equiv z_{m}(1) $~-- with $a$ suf\/f\/iciently
close to \textit{unity}, the $N$ initial coordinates $z_{m}(0) $
all suf\/f\/iciently well separated among themselves, and each $\tilde{z}%
_{m}(0) \equiv z_{m}(1) $ suf\/f\/iciently close to the
corresponding $z_{m}(0) $, see~(\ref{Expepsi})~-- which cause the
evolution yielded by the \textit{discrete-time} goldf\/ish model (\ref{DiscreteGold}) to \textit{mimic closely} that yielded by the \textit{continuous-time} goldf\/ish model~(\ref{ContModelalphaEq1}), provided at every
step of the \textit{discrete-time} evolution the appropriate identif\/ication
is made of the value of each twice-updated coordinate $\widetilde{\tilde{z}}_{n}\equiv z_{n}( \ell +2) $ (among the \textit{unordered set} of~$N$ values yielded by the \textit{discrete-time} equations of motion) by an
argument of \textit{contiguity} with $\tilde{z}_{n}\equiv z_{n} ( \ell
+1 ) $ and $z_{n}\equiv z_{n}( \ell ) $; and likewise an
appropriate identif\/ication is made by \textit{contiguity} of each coordinate
$z_{n} ( \ell +1 ) $ with the corresponding coordinate $z_{n} (
\ell  ) $ (among the \textit{unordered set} of $N$ values yielded by
Proposition~\ref{proposition2.1.1})~-- these arguments of \textit{contiguity} taking
the place of the \textit{continuity} of $z_{n}(t) $ as function
of $t$ applicable in the \textit{continuous-time} case. But the \textit{contiguity} argument breaks down if the positions at time~$\ell $ of two
dif\/ferent points, $z_{n}( \ell ) $ and $z_{m}( \ell ) $
with $n\neq m$, get too close to each other, corresponding to a
quasi-collision, or even coincide, corresponding to an actual collision;
which is however not featured by the \textit{generic} solution of the
\textit{discrete-time} model (\ref{DiscreteGold})~-- nor of the standard
goldf\/ish models~(\ref{NewtGoldb}) or~(\ref{GenGoldfish})~-- clearly emerging
only for a set of initial conditions $z_{n}(0)$, $\tilde{z}(
0) \equiv z_{n}(1) $ having \textit{unit} codimension in
the $2N$-dimensional (complex) phase space~$\underline{z}$,~$\underline{\tilde{z}}$.
\end{remark}

This \textit{important} remark is applicable to all the \textit{discrete-time} models considered below, although it will not be repeated.

%\textit{Remark 2.1.4}.
\begin{remark}\label{remark2.1.4}
Several of the formulas written above (in this
section) simplify somewhat via the following replacement of the
dependent variables:
\begin{gather*}
z_{n}( \ell ) \ \Rightarrow \ a^{\ell } z_{n}( \ell ),\qquad n=1,\dots,N.  %\label{Resc}
\end{gather*}
In particular the 3 equivalent versions (\ref{DiscreteGoldb}), (\ref{Equivb}) and (\ref{EqMotMod1ter}) of the discrete time equations of motion are
thereby reformulated to read
%\begin{subequations}
\begin{gather*}
\prod\limits_{j=1}^{N}\left( \frac{\widetilde{\tilde{z}}_{n}-z_{j}}{\widetilde{\tilde{z}}_{n}-\tilde{z}_{j}}\right) =1+a,\qquad n=1,\dots,N,
\\
\prod\limits_{j=1}^{N}\left( \frac{\widetilde{\tilde{z}}_{j}-z_{n}}{\tilde{z}
_{j}-z_{n}}\right) =\frac{1+a}{a},\qquad n=1,\dots,N,
\\
\prod\limits_{j=1}^{N}\left( \frac{\widetilde{\tilde{z}}_{j}-\tilde{z}_{n}}{z_{j}-\tilde{z}_{n}}\right) =-\frac{1}{a},\qquad n=1,\dots,N,
\end{gather*}
%\end{subequations}
and correspondingly the formula (\ref{SolMod1}) providing the solution of
the initial-value problem reads
\begin{gather*}
\prod\limits_{k=1}^{N}\left[ \frac{z_{n}( \ell )
-a^{-1} z_{k}(1) }{z_{n}( \ell ) -z_{k} (
0 ) }\right] =\frac{a^{\ell -1}-1}{a^{\ell }-1},\qquad n=1,\dots,N.
\end{gather*}
\end{remark}

Somewhat analogous remarks are applicable to all the \textit{discrete-time}
models considered below; their explicit implementation is left to the
interested reader.

\subsection{Second model}\label{section2.2}

In this subsection  we treat rather tersely a \textit{discrete-time}
dynamical system that generalizes the \textit{discrete-time} goldf\/ish model
described in the preceding Subsection~\ref{section2.1}. This generalization amounts to
the presence of an additional free parameter, $b$: indeed, for $b=0$ one
reobtains the model treated in the preceding Subsection~\ref{section2.1} (hence in this
subsection we assume that~$b$ does not vanish, $b\neq 0$).

The three equivalent versions of the equations of motion of this model read
as follows. The f\/irst version identif\/ies the twice updated coordinates $%
z_{n}( \ell +2) $ as the $N$ solution of the following equation
in $z$ (amounting to the identif\/ication of the $N$ roots of a polynomial of
degree $N$ in this variable):
\begin{subequations}
\label{EqMod2}
\begin{gather}
\sum_{k=1}^{N}\left\{ \left( \frac{\tilde{z}_{k}-az_{k}}{z-a\tilde{z}_{k}}
\right) \left( \frac{1+b\tilde{z}_{k}}{1+bz_{k}}\right)
 \prod\limits_{j=1,\; j\neq k}^{N}\left[ \left( \frac{\tilde{z}
_{k}-az_{j}}{\tilde{z}_{k}-\tilde{z}_{j}}\right) \left( \frac{1+b\tilde{z}
_{j}/a}{1+bz_{j}}\right) \right] \right\} =1.   \label{EqGenMod}
\end{gather}
The second and third versions consist of the following two systems:
\begin{gather}
 \sum_{k=1}^{N}\left[ \left( \frac{\widetilde{\tilde{z}}_{k}-a\tilde{z}_{k}
}{\tilde{z}_{k}-az_{n}}\right) \left( \frac{1+bz_{n}}{1+b\tilde{z}_{k}}
\right) \prod\limits_{j=1,\;j\neq k}^{N}\left( \frac{\widetilde{\tilde{z}}
_{j}-a\tilde{z}_{k}}{\tilde{z}_{j}-\tilde{z}_{k}}\right) \right] =a^{N-1},
\qquad
n=1,\dots,N;
\\
\left[ \prod\limits_{j=1}^{N}\left( \frac{\widetilde{\tilde{z}}_{j}-a
\tilde{z}_{n}}{az_{j}-\tilde{z}_{n}}\right) \right] \left[
\prod\limits_{j=1,\;j\neq n}^{N}\left( \frac{1+bz_{j}}{1+b\tilde{z}_{j}/a}
\right) \right] =a^{N-1}\frac{( 1+b\tilde{z}_{n}) }{(
1+bz_{n}) },  \qquad
n=1,\dots,N.
\end{gather}
\end{subequations}

\begin{remark}\label{remark2.2.1}
%\textit{Remark 2.2.1}.
This model~-- as the original goldf\/ish model (\ref{NewtGoldb}), and as the model treated above, see Remark \ref{remark2.1.1}~-- is \textit{invariant} under a rescaling of the dependent variables, $z_{n}\Rightarrow cz_{n}$ with $c$ an \textit{arbitrary constant}; but only
provided the parameter $b$ is also rescaled, $b\Rightarrow b/c$.
\end{remark}

The solution of this model is provided by an analog of (the f\/irst part of)
Proposition~\ref{remark2.1.1}, reading as follows:

%\textit{Proposition 2.2.1. }
\begin{proposition}\label{proposition2.2.1}
The $N$ values $z_{n}( \ell ) $ of
the dependent variables at the discrete time $\ell $ are the $N$ eigenvalues of the $N\times N$ matrix
\begin{subequations}
\label{SolUMod2}
\begin{gather}
U( \ell ) =U(0)  [ aI+bV(0)
 ] ^{\ell }  +V(0)  [  ( a-1 ) I+bV(0)  ]
^{-1}\big\{  [ aI+bV(0)  ] ^{\ell }-I\big\} ,
\label{GenSolU}
\end{gather}
where again $($see \eqref{U}$)$
\begin{gather}
U(0) =\text{\rm diag} [ z_{n}(0)  ]
,\qquad U_{nm}(0) =\delta _{nm}z_{n}(0),
\end{gather}
while the $N\times N$ matrix $V(0) $ is now defined
componentwise as follows:
\begin{gather}
 [ V(0)  ] _{nm}=\frac{v_{m}(0) }{1+bz_{m}(0) },\qquad n,m=1,\dots,N,  \label{A}
\end{gather}
\end{subequations}
with the quantities $v_{m}(0) $ defined again as in Subsection~{\rm \ref{section2.1} (}see \eqref{y} and the sentence following this formula$)$. Let us recall
that $I$ is the $N\times N$ unit matrix $($whose presence in \eqref{GenSolU}, however, might well be considered pleonastic$)$.
\end{proposition}

As evidenced by a comparison of (\ref{GenSolU}) with (\ref{UnmMod1}), the
behavior of the solutions of this model (with $b\neq 0$) are less simple
than those of the model discussed in the preceding Subsection~\ref{section2.1}. In
particular a \textit{confined} behavior emerges only, see (\ref{GenSolU}),
from initial data $z_{n}(0)$, $\tilde{z}_{n}(0)
\equiv z_{n}(1) $ implying, via~(\ref{A}) and~(\ref{y}), that
\textit{all} the $N$ eigenvalues of the $N\times N$ matrix $aI+bV(
0) $ have modulus not larger than \textit{unity}, reading $\exp
(-q_{n}+2\pi ir_{n})$ with the numbers~$q_{n}$ and~$r_{n}$ \textit{real}
and the $N$ numbers~$q_{n}$ \textit{nonnegative}, $q_{n}\geq 0$. If moreover
the $N$ numbers~$q_{n}$ \textit{all vanish} and the $N$ numbers $r_{n}$ are
\textit{all rational}, the behavior is \textit{periodic} (but not \textit{isochronous}, since these numbers, $q_{n}$ and $r_{n}$, generally depend on
the initial data; see (\ref{A}) and (\ref{y})). While, if some of (but not
all) the $N$ numbers $q_{n}$ are \textit{positive}, and none is \textit{negative}, then the phenomenology we just described (corresponding to the $q_{n}$'s \textit{all} vanishing) emerges only \textit{asymptotically}, as $\ell \rightarrow \infty $, up to corrections of order $\exp ( -q\ell
) $ with $q$ the smallest of the nonvanishing numbers $q_{n}$~-- provided all those~$r_{n}$'s are \textit{rational} whose corresponding $q_{n}$ \textit{vanish}.

Let us f\/inally mention that, also for this second model, a transition from
the \textit{discrete-time} independent variable $\ell $ to the \textit{continuous-time} variable $t$ can be performed (as tersely outlined at the
end of Subsection~\ref{section3.2}); but the \textit{continuous-time} goldf\/ish-type model
obtained in this manner turned out to be, to the best of our knowledge,
\textit{new}, hence it seemed appropriate to devote a~separate paper to it,
see~\cite{C2011}.

\subsection{Third model}\label{section2.3}

The third model is another one-parameter extension of the model treated in
Subsection~\ref{section2.1} (dif\/ferent from that treated in the preceding Subsection~\ref{section2.2}). Again its \textit{discrete-time} equations of motion can be presented
in three equivalent versions.

The f\/irst is characterized by this prescription: the twice-updated $N$
coordinates $\widetilde{\tilde{z}}_{n}\equiv z_{n}( \ell +2) $
are the $N$ roots of the following equation in the variable $z$,
\begin{subequations}
\label{Model3}
\begin{gather}
\sum_{k=1}^{N}\left[ \left( \frac{\tilde{z}_{k}-a_{+}z_{k}}{z-a_{+}\tilde{z}_{k}}\right) \prod\limits_{j=1,\; j\neq k}^{N}\left( \frac{\tilde{z}
_{k}-a_{+}z_{j}}{\tilde{z}_{k}-\tilde{z}_{j}}\right) \right] =\frac{1}{a_{-}},  \label{EvEq1}
\end{gather}
amounting again to the determination of the $N$ roots of a polynomial of
degree~$N$ in the variable~$z$. Here and below~$a_{+}$ and~$a_{-}$ are 2
arbitrary constants.

%\textit{Remark 2.3.1. }
\begin{remark}\label{remark2.3.1}
As entailed by a comparison of these \textit{discrete-time} second-order equations of motion with those of the f\/irst
model, see~(\ref{DiscreteGolda}), this third model coincides, for $a_{-}=1$,
with the f\/irst model with $a=a_{+}$.
\end{remark}

A neater formulation of these equations of motion reads (after
multiplication by~$a_{+}$, via~(\ref{Identity3b}) with $\zeta
_{k}=a_{+}^{2}z_{k}$, $\eta _{k}=a_{+}\tilde{z}_{k}$) as follows:
\begin{gather}
\prod\limits_{j=1}^{N}\left( \frac{z-a_{+}^{2}z_{j}}{z-a_{+}\tilde{z}_{j}}\right) =\frac{a_{-}+a_{+}}{a_{-}}.  \label{EvEq1a}
\end{gather}
\end{subequations}

An \textit{equivalent}, second formulation of this model is provided by the
following system of~$N$ polynomial equations for the twice-updated
coordinates $\widetilde{\tilde{z}}_{n}\equiv z_{n}( \ell +2) $:
\begin{subequations}
\label{EvEq22}
\begin{gather}
\sum_{k=1}^{N}\left[ \left( \frac{\widetilde{\tilde{z}}_{k}-a_{+}\tilde{z}
_{k}}{\tilde{z}_{k}-a_{+}z_{n}}\right) \prod\limits_{j=1,\; j\neq
k}^{N}\left( \frac{a_{+}\tilde{z}_{k}-\widetilde{\tilde{z}}_{j}}{\tilde{z}
_{k}-\tilde{z}_{j}}\right) \right] =a_{-}a_{+}^{N-1}, \qquad n=1,\dots,N;
\label{EvEq2a}
\end{gather}
and a neater version of these equations of motion reads (again via (\ref{Identity3b}), but now with $\zeta _{k}=\widetilde{\tilde{z}}_{k}$, $\eta
_{k}=a_{+}\tilde{z}_{k}$ and $z$ replaced by $a_{+}^{2}z_{n}$)
\begin{gather}
\prod\limits_{j=1}^{N}\left( \frac{\widetilde{\tilde{z}}_{j}-a_{+}^{2}z_{n}}{\tilde{z}_{j}-a_{+}z_{n}}\right) = ( a_{+}+a_{-} )
a_{+}^{N-1}, \qquad n=1,\dots,N.  \label{EvEq1b}
\end{gather}
\end{subequations}

And a third, also \textit{equivalent}, version of these equations of motion
reads as follows:
\begin{gather}
\prod\limits_{j=1}^{N}\left( \frac{\widetilde{\tilde{z}}_{j}-a_{+}\tilde{z}
_{n}}{a_{+}z_{j}-\tilde{z}_{n}}\right) =-a_{-}a_{+}^{N-1},\qquad n=1,\dots,N.
\label{EvEq33}
\end{gather}

\begin{remark}\label{remark2.3.2}
%\textit{Remark 2.3.2}.
Remark~\ref{remark2.1.1} also holds for this model.
\end{remark}

The solution of the initial-value problem for this \textit{discrete-time}
dynamical system is provided by the following

%\textit{Proposition 2.3.1}:
\begin{proposition}\label{proposition2.3.1}
The $N$ coordinates $z_{n}( \ell ) $
are the $N$ eigenvalues of the $N\times N$ matrix
\begin{subequations}
\label{SolU1}
\begin{gather}
U( \ell ) = ( a_{+} ) ^{\ell }C_{+}+ ( a_{-} )
^{\ell }C_{-},  \label{Usol3}
\end{gather}%
where the two constant $($i.e., $\ell $-independent$)$ $N\times N$ matrices $C_{+}$ and $C_{-}$ are defined in terms of the $2N$ initial data $z_{n}(0) $ and $\tilde{z}(0) \equiv z_{n}(1) $ by the formula
\begin{gather}
C_{\pm }=\pm ( a_{+}-a_{-}) ^{-1}[ U(1) -a_{\mp
}U(0) ]  \label{C+-1}
\end{gather}
with the two matrices $U(0) $ and $U(1) $ defined
componentwise as follows:
\begin{gather}
 [ U(0)  ] _{nm}=\delta _{nm}z_{n}(0) ,
\label{Uzero3}
\\
 [ U(1)  ] _{nm}=a_{+}^{1-N}\left[ \sum_{k=1}^{N}z_{k}
(1) -a_{+}\sum_{k=1,\; k\neq n}^{N}z_{k}(0) \right]
   \notag \\
\phantom{[ U(1)  ] _{nm}=}{}
\times \prod\limits_{j=1,\; j\neq m}^{N}\left[ \frac{a_{+}z_{m}(0)
-z_{j}(1) }{z_{m}(0) -z_{j}(0) }\right],\qquad n,m=1,\dots,N.  \label{Uone3}
\end{gather}
\end{subequations}

Note that we are, for simplicity, assuming that the two coupling constants $a_{\pm }$ are different, $a_{+}\neq a_{-}$ $($see \eqref{C+-1}$)$.
\end{proposition}

It is plain from these formulas that, if the two ``coupling constants'' $a_{\pm }$ (are dif\/ferent and) are conveniently written as follows,
\begin{gather}
a_{\pm }=\exp  ( -q_{\pm }+2\pi ir_{\pm } )  \label{aplusminus3}
\end{gather}%
with $q_{\pm }$ and $r_{\pm }$ \textit{real}, then, if the two numbers $%
q_{\pm }$ are both \textit{nonnegative}, $q_{\pm }\geq 0$, the time
evolution of the $N\times N$ matrix $U( \ell ) $ is bounded for
all values of the \textit{discrete-time} independent variable $\ell
=0,1,2,\dots$, hence its $N$ eigenvalues $z_{n}( \ell ) $ are all
as well bounded (the motion is conf\/ined); if in particular the two numbers $%
q_{\pm }$ both vanish, $q_{\pm }=0$, and the two numbers $r_{\pm }$ are both
\textit{rational} numbers, $r_{\pm }=K_{\pm }/L_{\pm }$ with $K_{+}$, $L_{+}$
and $K_{-}$, $L_{-}$ \textit{coprime integers} (and, for def\/initeness, $L_{\pm
}>0$), then the \textit{discrete-time} evolution of the matrix $U ( \ell
 ) $ is \textit{periodic} (with a~period~$L$ independent of the initial
data, being the minimum common multiple of $L_{+}$ and $L_{-}$, $L={\rm mcm} [ L_{+},L_{-} ] $) hence the \textit{discrete-time} dynamical
system~(\ref{Model3}) is \textit{isochronous}; while if, of the two numbers $
q_{\pm }$, one vanishes and the other is positive, $q_{+}=0$, $q_{-}=q>0$
respectively $q_{-}=0$, $q_{+}=q>0$, and $r_{+}$ respectively $r_{-}$ are
\textit{rational} numbers, then the \textit{isochronous} behavior (with
period $L_{+}$ respectively $L_{-}$) only emerges \textit{asymptotically},
as $\ell \rightarrow \infty $, up to corrections of order $\exp  (
-q\ell  ) $. While clearly if $q_{+}$ and $q_{-}$ are both \textit{positive} entailing (see~(\ref{aplusminus3})) $\left\vert a_{\pm
}\right\vert <0$, then (see (\ref{Usol3})) the matrix $U( \ell )
, $ hence as well all it eigenvalues $z_{n}(t) $, \textit{vanish}
asymptotically (as $t\rightarrow \infty $): $z_{n}( \infty ) =0$,
$n=1,\dots,N$.

Finally let us mention the transition from this \textit{discrete-time} model
to its \textit{continuous-time} counterpart. The treatment is completely
analogous to that detailed at the end of Subsection~\ref{section2.1}; except that now (\ref{aepsi}) must be replaced by
%\begin{subequations}
\begin{gather*}
a_{\pm } \ \Longrightarrow \ 1-i\omega _{\pm }\varepsilon
\end{gather*}
with
\begin{gather*}
\omega _{+}=\alpha \omega,\qquad \omega _{-}= ( \alpha -1 )\omega.
\end{gather*}
%\end{subequations}
It is then easily seen that again, at order $\varepsilon ^{0}=1$, one gets
from~(\ref{EvEq1a}) or~(\ref{EvEq1b}) or~(\ref{EvEq33}) a~trivial identity,
while at order $\varepsilon $ one gets the \textit{continuous-time} goldf\/ish
equations of motion~(\ref{NewtGoldb}); and the solution of this model, see Proposition~\ref{proposition2.3.1}, reproduces in this \textit{continuous-time}
limit the prescription~(\ref{Solz1}).

\subsection{Fourth model}\label{section2.4}

The fourth model is also characterized by three equivalent versions of its
\textit{discrete-time} equations of motion. The f\/irst consists of the
following prescription: the twice-updated $N$ coordinates $\widetilde{\tilde{z}}_{n}\equiv z_{n}( \ell +2) $ are the $N$ roots of the
following equation in the variable $z$,
\begin{subequations}
\label{ghat4}
\begin{gather}
\sum_{k=1}^{N}\left[ \frac{\hat{g}_{k} ( \underline{z},\underline{\tilde{%
z}} ) }{z-a\tilde{z}_{k}-b}\right] =\frac{1}{\gamma },  \label{EvEq2}
\end{gather}
where the $N$ quantities $\hat{g}_{k} ( \underline{z},\underline{\tilde{z}} ) $ are def\/ined as follows:
\begin{gather}
\hat{g}_{n} ( \underline{z},\underline{\tilde{z}} ) =a^{1-N}  (
\eta  \tilde{z}_{n}+\beta  )  \left[ \prod\limits_{j=1}^{N}\left( \frac{\tilde{z}_{n}-az_{j}-b}{\eta z_{j}+\beta }\right) \right]   \notag
\\
\phantom{\hat{g}_{n} ( \underline{z},\underline{\tilde{z}} ) =}{}
\times \left[ \prod\limits_{j=1,\; j\neq n}^{N}\left( \frac{\eta \tilde{z}
_{j}+a\beta -b\eta }{\tilde{z}_{n}-\tilde{z}_{j}}\right) \right]
,\qquad n=1,\dots,N.   \label{gnhat4}
\end{gather}
Throughout this subsection, the following Subsection~\ref{section3.4}, and Appendix~\ref{appendixB},
\begin{gather}
a=\alpha +\frac{\eta \rho }{1-\gamma },\qquad b=\frac{\beta\rho }{1-\gamma },  \label{ab}
\end{gather}
entailing
\begin{gather}
\alpha \beta =a\beta -b\eta, \qquad \eta\rho = ( a-\alpha  )
 ( 1-\gamma  ) ,\qquad \beta \rho =b ( 1-\gamma  ),
\end{gather}
\end{subequations}
where $\alpha $, $\beta$, $\gamma $, $\eta $ and $\rho $ are 5 arbitrary
constants (but the 3 constants $\beta $, $\eta $, $\rho $ only enter as $\beta \rho $ and $\eta \rho $, hence any one of these three constants could
be replaced by unity without signif\/icant loss of generality); in the
following we use interchangeably these constants in order to simplify some
formulas.

An \textit{equivalent} formulation of these \textit{discrete-time} equations
of motion reads as follows:
\begin{subequations}
\label{EqMot42}
\begin{gather}
\sum_{k=1}^{N}\left[ \frac{\check{g}_{k}\big( \underline{\tilde{z}},
\underline{\widetilde{\tilde{z}}}\big) }{ ( \eta \tilde{z}_{k}+\beta
 )  ( \tilde{z}_{k}-az_{n}-b ) }\right] =\frac{1}{ (
\eta z_{n}+\beta  ) },\qquad n=1,\dots,N,
\end{gather}
with
\begin{gather}
 \check{g}_{n}\big( \underline{\tilde{z}},\underline{\widetilde{\tilde{z}}}
\big) =\frac{a^{1-N}}{\gamma }\big( \widetilde{\tilde{z}}_{n}-a\tilde{z}
_{n}-b\big) \prod\limits_{j=1,\; j\neq k}^{N}\left( \frac{\widetilde{\tilde{
z}}_{j}-a\tilde{z}_{n}-b}{\tilde{z}_{j}-\tilde{z}_{n}}\right) ,  \qquad
n=1,\dots,N.  \label{gkMod42}
\end{gather}
\end{subequations}
It is a matter of trivial algebra to rewrite these equations of motion, (\ref{EqMot42}), as follows:
\begin{subequations}
\begin{gather}
 \sum_{k=1}^{N}\left[ \frac{1}{ ( \eta \tilde{z}_{k}+\beta  )
 ( \tilde{z}_{k}-az_{n}-b ) }\frac{\prod\limits_{j=1}^{N}\big(
\widetilde{\tilde{z}}_{j}-a\tilde{z}_{k}-b\big) }{\prod\limits_{j=1,\;j\neq k}^{N}\big( \tilde{z}_{j}-\tilde{z}_{k}\big) }\right] =\frac{\gamma
a^{N-1}}{\eta z_{n}+\beta },  \qquad
n =1,\dots,N.  \label{EqMotion4bis}
\end{gather}
And, as shown at the end of Appendix~\ref{appendixB}, a neater version of this system of
equations of motion then reads as follows:
\begin{gather}
\prod\limits_{j=1}^{N}\left[ \frac{\widetilde{\tilde{z}}
_{j}-a^{2}z_{n}-b ( 1+a ) }{\tilde{z}_{j}-az_{n}-b}\right]
-\prod\limits_{j=1}^{N}\left( \frac{\eta \widetilde{\tilde{z}}_{j}+\alpha
\beta }{\eta \tilde{z}_{k}+\beta }\right)
=\gamma a^{N-1}\frac{\eta az_{n}+\beta +\eta b}{\eta z_{n}+\beta },\nonumber\\
 n=1,\dots,N.  \label{EqEqMot4}
\end{gather}
\end{subequations}

And a third, \textit{equivalent} formulation of these equations of motion
reads as follows:
\begin{gather}
\check{g}_{n}\big( \underline{\tilde{z}},\underline{\widetilde{\tilde{z}}}
\big) =\hat{g}_{n} ( \underline{z},\underline{\tilde{z}} )
,\qquad n=1,\dots,N,  \label{Eqgg}
\end{gather}
with $\check{g}_{n}\big( \underline{\tilde{z}},\underline{\widetilde{\tilde{
z}}}\big) $ respectively $\hat{g}_{n} ( \underline{z},\underline{%
\tilde{z}} ) $ def\/ined by (\ref{gkMod42}) respectively (\ref{gnhat4}).

\begin{remark}\label{remark2.4.1}
%\textit{Remark 2.4.1}.
Above and below we assume for simplicity that the
parameters characterizing this model have \textit{generic} values, for
instance $\gamma \neq 0$ and $\gamma \neq 1$ (see~(\ref{EvEq2}) and~(\ref{ab})) and $a\neq 0$ (see (\ref{gkMod42})).
\end{remark}

%\textit{Remark 2.4.2}.
\begin{remark}\label{remark2.4.2}
This model~-- as the original goldf\/ish model (\ref{NewtGoldb}), and as the models treated above, see Remarks ~\ref{remark2.1.1},
\ref{remark2.2.1} and \ref{remark2.3.1}~-- is \textit{invariant} under a
rescaling of the dependent variables, $z_{n}\Rightarrow cz_{n}$ with $c$ an
\textit{arbitrary constant}; but only provided the parameter~$\beta $~-- hence as well the parameter $b$, see~(\ref{ab})~-- is also rescaled,
$\beta \Rightarrow c\beta$, $b\Rightarrow c b$.
\end{remark}

The solution of the initial-value problem for this \textit{discrete-time}
dynamical system is provided by the following

%\textit{Proposition 2.4.1}:
\begin{proposition}\label{proposition2.4.1}
The $N$ coordinates $z_{n}( \ell ) $
are the $N$ eigenvalues of the $N\times N$ matrix
\begin{subequations}
\label{Usol4}
\begin{gather}
U( \ell ) =U(0) P( 0,\ell -1)
+\sum_{k=1}^{\ell }\big[ ( B\gamma ^{k-1}+b) P( k,\ell
-1) \big],  \label{Umodel4}
\end{gather}
where the $N\times N$ matrix $P ( \ell _{1},\ell _{2} ) $ is
defined as follows%
\begin{gather}
P ( \ell _{1},\ell _{2} ) =\prod\limits_{j=\ell _{1}}^{\ell
_{2}} ( A\gamma ^{j}+a ),  \label{P}
\end{gather}%
and the two $\ell $-independent $N\times N$ matrices $A$ and $B$ are defined
as follows%
\begin{gather}
A=\eta V(0) -\frac{\eta }{\beta }b,\qquad B=\beta V(0) -b.  \label{AB}
\end{gather}
\end{subequations}
Here and throughout we use the convention that $($for arbitrary finite $X_{j})$
$\prod\limits_{j=\ell _{1}}^{\ell _{2}}X_{j}=I$ if $\ell _{1}>\ell _{2}$
and $\sum\limits_{k=k_{1}}^{k_{2}}X_{j}=0$ if $k_{1}>k_{2}$. As for the
two $N\times N$ matrices $U(0) $ and $V(0) $, they
are defined in terms of the $2N$ initial data $z_{n}(0) $ and $%
\tilde{z}_{n}(0) \equiv z_{n}(1) $ as follows:
%\begin{subequations}
\begin{gather*}
U(0) =Z(0) =\text{\rm diag} [ z_{n}(0)
 ],  \label{U(0)2}
\\
V(0) = [ \eta Z(0) +\beta  ] ^{-1}\big\{
M(0) Z(1)  [ M(0)  ]
^{-1}-\alpha Z(0) \big\} ,  \label{V(0)2}
\end{gather*}
with the $N\times N$ matrices $Z( \ell ) $ and $M(0) $
defined, componentwise, as follows:
\begin{gather*}
Z( \ell ) =\text{\rm diag}[ z_{n}( \ell ) ];
\qquad Z_{nm}( \ell ) =\delta _{nm}z_{n}( \ell )
,\qquad n,m=1,\dots,N,
\\
M_{nm}(0) =\frac{\hat{g}_{m}(0) }{z_{m}(
1) -az_{n}(0) -b},\qquad n,m=1,\dots,N,
\end{gather*}
where the notation $\hat{g}_{m}(0) $ is an abbreviation for $
\hat{g}_{m} ( \underline{z},\underline{\tilde{z}} ) $, see \eqref{gnhat4}, evaluated at $\underline{z}=\underline{z}(0)$,
$\underline{\tilde{z}}=\underline{\tilde{z}}(0) \equiv \underline{z}(1) $. Note that  Lemma~{\rm \ref{lemmaA5}} $($with $f_{n}=1$, $g_{m}=\hat{g}_{m}(0) $, $\xi _{m}=z_{m}(1) $, $\eta _{n}=az_{n}( 0) +b)$ entails the following componentwise definition
of the inverse matrix $[ M(0) ] ^{-1}$:
\begin{gather*}
 \big\{  [ M(0)  ] ^{-1}\big\} _{nm}=a^{1-N} \left[
\frac{z_{n}(1) -a z_{m}(0) -b}{\hat{g}_{n}(
0) }\right]   \left\{ \prod\limits_{j=1,\; j\neq n}^{N}\left[ \frac{z_{j}(
1) -az_{m}(0) -b}{z_{j}(1) -z_{n}(
1) }\right] \right\} \nonumber\\
\phantom{\big\{ [ M(0) ] ^{-1}\big\} _{nm}=}{}\times
\left\{ \prod\limits_{j=1,\; j\neq m}^{N}\left[
\frac{z_{n}(1) -az_{j}(0) -b}{z_{m}(0)
-z_{j}(0) }\right] \right\},  \qquad n,m =1,\dots,N,
\end{gather*}
hence an explicit expression of the $N\times N$ matrix $V(0) $
reads, componentwise, as follows:
\begin{gather*}
V_{nm}(0) =-\frac{\alpha z_{n}(0) }{\eta
z_{n}(0) +\beta }\delta _{nm}
+\frac{a^{1-N}}{\eta z_{n}(0) +\beta }\sum_{k=1}^{N}\left[
z_{k}(1) \left\{ \prod\limits_{j=1,\; j\neq k}^{N}\left[ \frac{
z_{j}(1) -az_{m}(0) -b}{z_{j}(1)
-z_{k}(1) }\right] \right\} \right.  \notag \\
\left. \phantom{V_{nm}(0) =}{}
\times \left\{ \frac{\prod\limits_{j=1,\; j\neq n}^{N}\left[
z_{k}(1) -az_{j}(0) -b\right] }{\prod\limits_{j=1,\; j\neq m}\left[ z_{m}(0) -z_{j}(0)
\right] }\right\} \right]  .
\end{gather*}
%\end{subequations}
\end{proposition}

%\textit{Remark 2.4.3}.
\begin{remark}\label{remark2.4.3}
It is relevant to this expression, (\ref{Usol4}), of
the $N\times N$ matrix $U( \ell ) $~-- whose $N$ eigenvalues
provide the $N$ coordinates $z_{n}( \ell ) $~-- that (\ref{P}) and
(\ref{AB}) entail
%\begin{subequations}
\begin{gather*}
P ( \ell _{1},\ell _{2} ) =Q\, \text{diag} [ p_{n} ( \ell
_{1},\ell _{2} )  ] Q^{-1},
\qquad
A=Q\, \text{diag} [ a_{n} ] Q^{-1},\qquad B=Q\, \text{diag} [ b_{n} ] Q^{-1},
\\
p_{n} ( \ell _{1},\ell _{2} ) =\prod\limits_{j=\ell _{1}}^{\ell
_{2}} ( a_{n}\gamma ^{j}+a ) ,\qquad a_{n}=\eta v_{n}-\frac{\eta }{\beta }b,\qquad b_{n}=\beta v_{n}-b,  \qquad
n=1,\dots,N,
\end{gather*}
with $v_{n}$ the $N$ ($\ell $-independent) eigenvalues of the $N\times N$
matrix $V(0) $ and $Q$ the corresponding ($\ell $-independent)
diagonalizing matrix,
\begin{gather*}
V(0) =Q\,\text{diag} ( v_{n} )  Q^{-1}.
\end{gather*}
%\end{subequations}
\end{remark}

And let us mention that, also for this fourth model, a transition from the
\textit{discrete-time} independent variable $\ell $ to the \textit{continuous-time} variable $t$ can be performed (see the end of Subsection~\ref{section3.4}). And, as in the case of the second model, also in this case the \textit{continuous-time} goldf\/ish model thereby obtained turned out to be, to the
best of our knowledge, \textit{new}. Hence it seemed appropriate to devote
to this model a separate paper~\cite{C2011c}.

\section{Proofs}\label{section3}

In this section  we prove the f\/indings reported in the preceding Section~\ref{section2}.

The basic strategy to obtain all these results goes as follows. The starting
point is a \textit{solvable} system of two \textit{matrix} f\/irst-order
\textit{discrete-time} ODEs, say
\begin{gather}
\tilde{U}=F_{1}( U,V) ,\qquad \tilde{V}=F_{2}( U,V),
\label{UVODEs}
\end{gather}%
where $\ell =0,1,2,\dots$ is the \textit{discrete-time} independent variable,
the two dependent variables $U\equiv U( \ell )$, $V\equiv V(
\ell ) $ are $N\times N$ matrices and of course superimposed tildes
denote the updating of the \textit{discrete-time}, $\tilde{U}\equiv U(
\ell +1)$, $\tilde{V}\equiv V( \ell +1) $. The \textit{solvable} character of this matrix system entails the possibility to obtain
\textit{explicitly} the solution of its initial-value problem. Four cases
when this is possible~-- corresponding to~4 simple assignments of the
functions $F_{1}( U,V) $ and $F_{2}( U,V) $~-- are
treated in the following 4 subsections. Note that the two functions $F_{1}(U,V)$, $F_{2}(U,V)$ are assumed to depend on no other matrix besides~$U$
and~$V$ (and the unit matrix~$I$); they may of course feature some \textit{scalar} constants, and the order in which the two, generally noncommuting,
matrices $U$ and $V$ appear in their def\/inition is of course relevant: see
below.

One assumes moreover that the $N\times N$ matrix $U\equiv U( \ell
) $ is \textit{diagonalizable} and denotes as $R\equiv R( \ell
) $ the diagonalizing $N\times N$ matrix:
\begin{subequations}
\label{UZ}
\begin{gather}
U\equiv RZR^{-1},\qquad U( \ell ) \equiv R( \ell )
Z( \ell ) [ R( \ell ) ] ^{-1},
\label{Udiag}
\\
Z=\text{diag}[ z_{n}] ,\qquad Z( \ell) =\text{diag}[
z_{n}( \ell ) ] ,  \label{Zdiag}
\end{gather}
\end{subequations}
where the notation $z_{n}( \ell ) $ for the $N$ eigenvalues of
the $N\times N$ matrix $U\equiv U( \ell ) $ shall be justif\/ied by
the identif\/ication, see below, of these quantities with the dependent
variables of the \textit{discrete-time} dynamical systems introduced above.

%\textit{Remark 3.1}.
\begin{remark}\label{remark3.1}
These formulas entail that the matrix $R( \ell
) $ is def\/ined up to right-multiplication by an arbitrary \textit{diagonal} matrix $D( \ell )$, $R( \ell ) \Rightarrow
R( \ell )  D( \ell ) $.
\end{remark}

Next we introduce the two matrices $M( \ell ) $ and $Y( \ell
) $ def\/ined as follows:
\begin{subequations}
\label{MVel}
\begin{gather}
M=R^{-1}\tilde{R},\qquad M( \ell ) =[ R( \ell )
] ^{-1}R( \ell +1),  \label{Mel}
\\
V=RY\tilde{R}^{-1},\qquad V( \ell ) =R( \ell ) Y(
\ell ) [ R( \ell +1) ] ^{-1},  \label{Vel}
\end{gather}
so that
\begin{gather}
V=RYM^{-1}R^{-1},\qquad V( \ell ) =R( \ell ) Y(
\ell ) [ M( \ell ) ] ^{-1}[ R( \ell
) ] ^{-1}.  \label{VMR}
\end{gather}
\end{subequations}

%\textit{Remark 3.2}.
\begin{remark}\label{remark3.2}
The element of freedom in the def\/inition of the matrix $%
R( \ell ) $, see  Remark~\ref{remark3.1}, entails that the matrix $M( \ell ) $ is def\/ined up to the ``gauge transformation'' resulting
by inserting in its def\/inition (\ref{Mel}) the $N\times N$ matrix $[
R( \ell ) D( \ell ) ] ^{-1}=[ D( \ell
) ] ^{-1}[ R( \ell ) ] ^{-1}$ in place of
the matrix $[ R( \ell ) ] ^{-1}$ (and of course $R( \ell +1) D( \ell +1) $ in place of $R( \ell
+1) $): hence, as a consequence of the \textit{arbitrary} nature of
the \textit{diagonal} matrix $D( \ell ) ,$ out of the $N^{2}$
elements of the $N\times N$ matrix $M\equiv M( \ell ) $ only $N^{2}-N$ are signif\/icant. Likewise for the matrix $Y\equiv Y( \ell
) $.
\end{remark}

One then, by inserting (\ref{Udiag}) and (\ref{VMR}) in (\ref{UVODEs}),
obtains the following system of two f\/irst-order \textit{discrete-time} $N\times N$ matrix evolution equations:
\begin{gather}
M \tilde{Z}=F_{1}\big( Z, Y M^{-1}\big)  M ,\qquad M \tilde{Y}=F_{2}\big(
Z, Y M^{-1}\big)  M \tilde{M};
 \label{EqZY}
\end{gather}
and from these two matrix equations, by making a convenient \textit{ansatz}
for the two matrices~$M\equiv M ( \ell  ) $ and $Y\equiv Y (
\ell  ) $ in terms of the $2N$ quantities $z_{n}\equiv z_{n} (\ell
 ) $ and $\tilde{z}_{n}\equiv z_{n} ( \ell +1 ) $~-- an \textit{ansatz} which must of course be consistent with these two matrix evolution
equations~-- one obtains a~system of $N$ second-order discrete-time evolution
equations for the $N$ coordinates $z_{n}\equiv z_{n} ( \ell  ) $.
This last step is of course only possible for \textit{special} assignments,
in the \textit{discrete-time} matrix evolution equations (\ref{UVODEs}), of
the two matrix functions $F_{1} ( U,V) $ and $F_{2}(U,V) $, see below.

The \textit{discrete-time} dynamical system thereby obtained is then \textit{solvable}, since the quantities $z_{n}\equiv z_{n}( \ell ) $ are
the $N$ eigenvalues of the $N\times N$ matrix $U\equiv U( \ell ) $
which, as solution of the, assumedly \textit{solvable}, matrix evolution
system~(\ref{UVODEs}), can be \textit{explicitly} evaluated. How this works
out is shown in detail in the following subsections: in more detail in
Subsection~\ref{section3.1}, where the simplest case is treated.

Let us also mention, once and for all, that in the following we will
conveniently assume that the matrix $U$ is \textit{initially} diagonal:
\begin{subequations}
\label{IniUR}
\begin{gather}
U( 0) =Z( 0) \equiv \text{diag}[ z_{n}(0) ] ,  \label{Uzero}
\end{gather}
implying (up to the ambiguity mentioned above, see Remark~\ref{remark3.1})
\begin{gather}
R( 0) =I.  \label{R(0)EqI}
\end{gather}
\end{subequations}
Here and throughout $I$ is the $N\times N$ \textit{unit} matrix, i.e.,
componentwise, $I_{nm}=\delta _{nm}$.

\subsection{Solution of the f\/irst model}\label{section3.1}

The point of departure to obtain the f\/indings reported in Subsection~\ref{section2.1} is
the following \textit{discrete-time} f\/irst-order, linear, matrix system (see~(\ref{UVODEs})):
\begin{subequations}
\begin{gather}
\tilde{U}=aU+V, \qquad \tilde{V}=V,  \label{EqU}
\end{gather}
where $a$ is an arbitrary scalar constant. Note that the second of these two
ODEs entails that in this case $V$ is a constant (i.e., $\ell $-independent) $N\times N$ matrix, $V( \ell) =V( 0) $.
It is plain that the solution of the corresponding initial-value problem for
the $N\times N$ matrix $U$ reads
\begin{gather}
U( \ell) =U( 0) a^{\ell }+V( 0) \frac{a^{\ell }-1}{a-1}.  \label{SolU}
\end{gather}
\end{subequations}

Let us now proceed as indicated in the f\/irst part of Section~\ref{section3}. It is then
easily seen (via~(\ref{UZ}) and~(\ref{MVel})) that the f\/irst of the two
\textit{discrete-time} matrix evolution equations (\ref{EqU}) yields (see~(\ref{EqZY})) the matrix equation
\begin{subequations}
\begin{gather}
M\tilde{Z}-aZM=Y,  \label{EqMZ}
\end{gather}
namely, componentwise,
\begin{gather}
M_{nm}=\frac{Y_{nm}}{\tilde{z}_{m}-az_{n}},\qquad n,m=1,\dots,N.  \label{MV}
\end{gather}
\end{subequations}

Likewise, the second of the two \textit{discrete-time} matrix evolution
equations (\ref{EqU}) yields the matrix relation
%\begin{subequations}
\begin{gather*}
Y\tilde{M}=M\tilde{Y}.  %\label{EqEM}
\end{gather*}

Via (\ref{MV}) this matrix equation implies the following $N^{2}$ relations:
\begin{gather}
\sum_{k=1}^{N}\left\{ Y_{nk}\tilde{Y}_{km}\left[ \big( \widetilde{\tilde{z}}_{m}-a\tilde{z}_{k}\big) ^{-1}-\left( \tilde{z}_{k}-az_{n}\right)
^{-1}\right] \right\} =0,  \qquad
 n,m=1,\dots,N.  \label{EqMV}
\end{gather}
%\end{subequations}

This derivation shows that this system of $N^{2}$ \textit{discrete-time}
equations of motion is equivalent to the \textit{solvable} equation of
motion (\ref{EqU}) for the $N\times N$ matrix $U$; hence it is just as
\textit{solvable}. Note that the dependent variables are now the $N$
coordinates $z_{n}$ and the $N^{2}$ matrix elements~$Y_{nm}$ (of which only $N( N-1) $ are signif\/icant, see Remark~\ref{remark3.2}; so the
number of equations and the number of dependent variables tally). To obtain
a model that qualif\/ies as \textit{discrete-time} analog of the \textit{%
continuous-time} goldf\/ish model~(\ref{ContGoldAlphaEqOne}) we need to
distill from this system a set of \textit{only} $N$ equations of motion
involving \textit{only} the $N$ coordinates~$z_{n}$. The standard trick to
do so (see, for instance, Section~4.2.2 entitled ``Goldf\/ishing'' of \cite{C2008}) is to identify~-- if possible~-- an \textit{ansatz} which expresses
the $N^{2}$ components of the matrix $Y$ in terms of the $2N$ quantities~$z_{n}$, $\tilde{z}_{n}$, yielding $N$ equations of motion involving \textit{only} the $N$ coordinates $z_{n}$, $\tilde{z}_{n}$ and $\widetilde{\tilde{z}}_{n}$~-- to be interpreted as equations of motion of the \textit{discrete-time} goldf\/ish~-- and implying that the $N^{2}$ equations of motion~(\ref{EqMV})
are all satisf\/ied, thanks to these very equations of motion.

An educated guess for such an \textit{ansatz} reads as follows:
\begin{gather}
Y_{nm}=g_{m},\qquad n,m=1,\dots,N.  \label{Vans}
\end{gather}%
Note that we reserve at this stage the option to assign the $N$ quantities $g_{m}$.

Via this \textit{ansatz} the equations (\ref{EqMV}) become
\begin{subequations}
\begin{gather}
\sum_{k=1}^{N}\left( \frac{g_{k}}{\widetilde{\tilde{z}}_{m}-a\tilde{z}_{k}}-
\frac{g_{k}}{\tilde{z}_{k}-a z_{n}}\right) =0,\qquad n,m=1,\dots,N,  \label{Cond}
\end{gather}%
hence they amount to the following 2 systems, each involving \textit{only} $N $ equations:
\begin{gather}
\sum_{k=1}^{N}\left( \frac{g_{k}}{\tilde{z}_{k}-az_{n}}\right)
=1,\qquad n=1,\dots,N,  \label{Conda}
\\
\sum_{k=1}^{N}\left( \frac{g_{k}}{\widetilde{\tilde{z}}_{n}-a\tilde{z}_{k}}\right) =1,\qquad n=1,\dots,N.  \label{Condb}
\end{gather}
\end{subequations}
The unit in the right-hand sides could of course be replaced by an arbitrary
constant~$c$~-- of course the same constant in (\ref{Conda}) and (\ref{Condb})~-- but this would merely entail an irrelevant rescaling of $g_{k}$ by $c$; see below.

These are now two sets of $N$ equations, each featuring \textit{linearly}
the $N$ quantities $g_{k},$ that we like to eliminate in order to obtain a
set of $N$ ``equations of motion'' determining the twice updated coordinates $%
\widetilde{\tilde{z}}_{n}\equiv z_{n}( \ell +2) $ in terms of the
$2N$ coordinates $z_{n}\equiv z_{n}( \ell ) $ and $\tilde{z}_{n}\equiv z_{n}( \ell +1)$. There are three alternative
strategies to achieve this goal. One can solve the f\/irst \textit{linear}
system thereby obtaining $g_{k}$ as a function of $\underline{z}$ and $\underline{\tilde{z}},$ and then insert this expression $g_{k}\equiv \hat{g}_{k}\left( \underline{z},\underline{\tilde{z}}\right) $ in the second
system; alternatively, one can solve the second \textit{linear} system,
thereby obtai\-ning~$g_{k}$ as a function of $\underline{\tilde{z}}$ and $\underline{\widetilde{\tilde{z}}}$, and then insert this expression $g_{k}\equiv \check{g}_{k}\big( \underline{\tilde{z}},\underline{\widetilde{\tilde{z}}}\big) $ in the f\/irst system; or one can equate the two
expressions of $g_{k}$ obtained solving the f\/irst, respectively the second,
system, i.e.\ write $\hat{g}_{k}( \underline{z},\underline{\tilde{z}}) =\check{g}_{k}\big( \underline{\tilde{z}},\underline{\widetilde{\tilde{z}}}\big) $. Clearly the three sets of equations of motion obtained
in this manner are \textit{equivalent}, i.e.\ they characterize the same
\textit{discrete-time} evolution of the $N$ coordinates $z_{n}\equiv
z_{n}( \ell ) $; but they may seem quite dif\/ferent (indeed, see (\ref{DiscreteGold}), (\ref{Equiv}) and (\ref{EqMotMod1ter})). Note that we
introduced a superimposed decoration on the functions $\hat{g}_{k}(\underline{z},\underline{\tilde{z}}) $ respectively $\check{g}_{k}\big( \underline{\tilde{z}},\underline{\widetilde{\tilde{z}}}\big) $
to emphasize that the functional dependence on their arguments is generally
dif\/ferent, as implied by their def\/initions as solutions of~(\ref{Conda})
respectively of~(\ref{Condb}).

Let us f\/irst of all see what the f\/irst approach yields. From~(\ref{Conda})
one obtains (via Lemma~\ref{lemmaA1} reported in   Appendix~\ref{appendixA}, with $\xi
_{k}=\tilde{z}_{k}$, $\eta _{n}=az_{n}$, $c=1$) the following expression of
$g_{k}\equiv \hat{g}_{k}( \underline{z},\underline{\tilde{z}}) $:
\begin{gather}
\hat{g}_{k}( \underline{z},\underline{\tilde{z}}) =( \tilde{z}_{k}-az_{k}) \prod\limits_{j=1,\; j\neq k}^{N}\left( \frac{\tilde{z}
_{k}-az_{j}}{\tilde{z}_{k}-\tilde{z}_{j}}\right) ,\qquad k=1,\dots,N.
\label{gk}
\end{gather}
The insertion of this expression of $g_{k}$ in~(\ref{Condb}) yields the
equations of motions~(\ref{DiscreteGolda}).

Likewise, the second approach yields, from (\ref{Condb}) (again via  Lemma~\ref{lemmaA1}, but now with $\xi _{k}=a\tilde{z}_{k}$, $\eta _{n}=\widetilde{\tilde{z}}_{n}$, $c=-1$) the following expression of $g_{k}\equiv \check{g}_{k}\big( \underline{\tilde{z}},\underline{\widetilde{\tilde{z}}}\big) $:
\begin{gather}
\check{g}_{k}\big( \underline{\tilde{z}},\underline{\widetilde{\tilde{z}}}
\big) =a^{1-N}\big( \widetilde{\tilde{z}}_{k}-a \tilde{z}_{k}\big)
 \prod\limits_{j=1,\; j\neq k}^{N}\left( \frac{\widetilde{\tilde{z}}_{j}-a\tilde{z}_{k}}{\tilde{z}_{j}-\tilde{z}_{k}}\right) ,\qquad k=1,\dots,N.
\label{gmbis}
\end{gather}
The second version, (\ref{Equiva}), of the \textit{discrete-time} equations
of motion follows by inserting this expression of~$g_{k}$ in~(\ref{Conda}).

And the third approach yields, by equating~(\ref{gk}) to (\ref{gmbis}), the
third version, (\ref{EqMotMod1ter}), of the \textit{discrete-time} equations
of motion.

We have seen that the solutions $z_{n} ( \ell  ) $ of these \textit{discrete-time} equations of motion are provided by the eigenvalues of the $N\times N$ matrix $U ( \ell  ) $, see~(\ref{SolU}). To prove
 Proposition~\ref{proposition2.1.1} we must now obtain from (\ref{SolU}) (also taking
advantage of the \textit{ansatz} (\ref{Vans})) the expression (\ref{U}) of
this matrix in terms of the initial data $z_{n} ( 0 ) $, $\tilde{z}_{n} ( 0 ) \equiv z_{n} ( 1 ) $ of the \textit{discrete-time} dynamical system.

This requires that we express the two matrices $U ( 0 ) $ and $%
V(0) $ appearing in the right-hand side of (\ref{SolU}) in terms
of the initial data $z_{n} ( 0 ) $, $\tilde{z}_{n} ( 0 )
\equiv z_{n} ( 1 ) $.

The expression of $U ( 0 ) $ is an immediate consequence of (\ref{Uzero}):
\begin{gather}
[ U( 0) ] _{nm}=\delta _{nm} z_{n} ( 0 ) .
\label{Uzeronm}
\end{gather}

To obtain $V( 0) $ we note f\/irst of all that (\ref{VMR}) and (\ref{R(0)EqI}) imply
%\begin{subequations}
\begin{gather*}
V( 0) =Y( 0) [ M( 0) ] ^{-1},
\end{gather*}
while (\ref{MV}) with the \textit{ansatz} (\ref{Vans}) implies (at $\ell =0$){\samepage
\begin{gather*}
M_{nm}( 0) =\frac{\hat{g}_{m}( 0) }{z_{m}(
1) -az_{n}( 0) },\qquad n,m=1,\dots,N,
\end{gather*}
where of course $\hat{g}_{m}( 0) $ stands for $\hat{g}_{m}(
\underline{z},\underline{\tilde{z}}) $, see~(\ref{gk}), evaluated at $\underline{z}=\underline{z}( 0) $, $\underline{\tilde{z}}=
\underline{\tilde{z}}( 0) \equiv \underline{z}( 1) $.}

We then evaluate the matrix $[ M( 0)] ^{-1}$ via
 Lemma \ref{lemmaA5} (with $\xi _{m}=z_{m}( 1) ,$ $\eta
_{n}=a~z_{n}( 0) $, $f_{n}=1$ and $g_{m}=\hat{g}_{m}(
0)$) and, using again the \textit{ansatz}~(\ref{Vans}) (at $\ell =0$), we obtain the following expression of the $N\times N$ matrix $V(0)$:
\begin{gather*}
[ V( 0)] _{nm}=v_{m}( 0)u_{nm},\qquad n,m=1,\dots,N,
\end{gather*}
with $v_{m}( 0) $ def\/ined as in Subsection~\ref{section2.1} (see~(\ref{y}) and
the sentence following this formula) and
\begin{gather*}
u_{nm}=\sum_{k=1}^{N}\left\{ \frac{\prod\limits_{j=1,\; j\neq m}^{N} [
z_{k} ( 1 ) -az_{j} ( 0 )  ] }{\prod\limits_{j=1,\;j\neq k}^{N} [ z_{k} ( 1 ) -z_{j} ( 1 )  ] }
\right\} ,\qquad n,m=1,\dots,N.  \label{u}
\end{gather*}%
But the identity (\ref{Identity3a}) (with $\eta _{k}=z_{k}( 1) $,
$\zeta _{j}=az_{j}( 0) $) entails $u_{nm}=1$, hence
\begin{gather}
[ V( 0)] _{nm}=v_{m}( 0) ,\qquad n,m=1,\dots,N.
\label{Vzeronm}
\end{gather}
%\end{subequations}

The insertion of these expressions of $U( 0) $ and $V(0)$, (\ref{Uzeronm}) and~(\ref{Vzeronm}), in~(\ref{SolU}) yields~(\ref{U}), thereby completing the proof of  Proposition~\ref{proposition2.1.1}.

Let us now provide a terse treatment of the transition from the \textit{discrete-time} equations of motion~(\ref{DiscreteGoldb}), which we
conveniently re-write here as follows,
\begin{gather}
\prod\limits_{j=1}^{N}\left( \frac{\widetilde{\tilde{z}}_{n}-a^{2}z_{j}}{\widetilde{\tilde{z}}_{n}-a~\tilde{z}_{j}}\right) =1+a,\qquad n=1,\dots,N,
\label{EqMod1}
\end{gather}
to the \textit{continuous-time} case, see~(\ref{ContGoldAlphaEqOne}). It is
then appropriate to treat separately the factor with $j=n$ in the product
appearing in the left-hand side of~(\ref{EqMod1}), and all the other factors
with $j\neq n$. The basic equations are~(\ref{Expepsi}), entailing
%\begin{subequations}
\begin{gather*}
\widetilde{\tilde{z}}_{n}-a^{2}z_{n}=2 \varepsilon ( \dot{z}_{n}+i\omega z_{n}) +\varepsilon ^{2}\big( 2\ddot{z}_{n}+\omega
^{2}z_{n}\big) +O\big( \epsilon ^{3}\big) ,
\\
\widetilde{\tilde{z}}_{n}-a\tilde{z}_{n}=\varepsilon ( \dot{z}
_{n}+i\omega z_{n}) +\frac{\varepsilon ^{2}}{2}( 3\ddot{z}
_{n}+2i\omega \dot{z}_{n}) +O\big( \epsilon ^{3}\big) ,
\\
\widetilde{\tilde{z}}_{n}-a^{2}z_{j}=z_{n}-z_{j}+2\varepsilon ( \dot{z}_{n}+i\omega z_{j}) +O\big( \varepsilon ^{2}\big) ,\qquad j\neq n,
\\
\widetilde{\tilde{z}}_{n}-a\tilde{z}_{j}=z_{n}-z_{j}+\varepsilon ( 2\dot{z}_{n}-\dot{z}_{j}+i\omega z_{j}) +O\big( \varepsilon
^{2}\big) ,\qquad j\neq n.
\end{gather*}
%\end{subequations}
Hence, after a little algebra,
\begin{subequations}
\label{lim1}
\begin{gather}
\frac{\widetilde{\tilde{z}}_{n}-a^{2}~z_{n}}{\widetilde{\tilde{z}}_{n}-a
\tilde{z}_{n}}=2+\varepsilon \frac{-\ddot{z}_{n}-2i\omega \dot{z}
_{n}+\omega ^{2}z_{n}}{\dot{z}_{n}+i\omega z_{n}}+O\big( \varepsilon
^{2}\big) ,
\\
\frac{\widetilde{\tilde{z}}_{n}-a^{2}z_{j}}{\widetilde{\tilde{z}}_{n}-a
\tilde{z}_{j}}=1+\varepsilon \frac{\dot{z}_{j}+i\omega z_{j}}{z_{n}-z_{j}}
+O\big( \varepsilon ^{2}\big) ,\qquad j\neq n,
\end{gather}%
implying
\begin{gather}
 \prod\limits_{j=1}^{N}\left( \frac{\widetilde{\tilde{z}}_{n}-a^{2}z_{j}}{\widetilde{\tilde{z}}_{n}-a\tilde{z}_{j}}\right) =\left( 2+\varepsilon
\frac{-\ddot{z}_{n}-2i\omega \dot{z}_{n}+\omega ^{2}z_{n}}{\dot{z}_{n}+i\omega z_{n}}\right)  \notag \\
\hphantom{\prod\limits_{j=1}^{N}\left( \frac{\widetilde{\tilde{z}}_{n}-a^{2}z_{j}}{\widetilde{\tilde{z}}_{n}-a\tilde{z}_{j}}\right) =}{}
\times  \left[ 1+\varepsilon \sum_{j=1,\; j\neq n}^{N}\left( \frac{\dot{z}
_{j}+i\omega z_{j}}{z_{n}-z_{j}}\right) \right] +O\big( \varepsilon
^{2}\big) .  \label{54a}
\end{gather}
While of course
\begin{gather}
1+a=2-i\omega \varepsilon ,  \label{54b}
\end{gather}
\end{subequations}
see (\ref{aepsi}). It is then clear that the insertion of these two
formulas, (\ref{54a}) and (\ref{54b}), in~(\ref{EqMod1}) yields, at order $\varepsilon ^{0}=1,$ the trivial identity $2=2$, and at order $\varepsilon $
the equations of motion of the \textit{continuous-time} goldf\/ish model (\ref{ContGoldAlphaEqOne}).

In an analogous manner one reobtains (\ref{ContGoldAlphaEqOne}) from (\ref{Equivb}) or from (\ref{EqMotMod1ter}).

Let us also show that (\ref{SolMod1}), which we rewrite here conveniently as
follows,
\begin{gather}
\prod\limits_{k=1}^{N}\left[ \frac{z-a^{\ell }z_{k}( 1) a^{-1}}{z-a^{\ell }z_{k}( 0) }\right] =\frac{a^{\ell }a^{-1}-1}{a^{\ell
}-1},  \label{SolMod1bis}
\end{gather}
yields, in the \textit{continuous-time} limit, (\ref{SolvContAlphaEqOne}).
Indeed the relation $a=1-i\varepsilon \omega $ (see (\ref{aepsi})) entails
%\begin{subequations}
\begin{gather*}
a^{-1}=1+i\varepsilon \omega +O\big( \varepsilon ^{2}\big) ,
\\
a^{\ell }=\exp (-i\omega t)\left( 1-\varepsilon \frac{\omega ^{2}t}{t}\right) +O\big( \varepsilon ^{2}\big)
\end{gather*}
(via the f\/irst of the three relations (\ref{landaepsi})), and
\begin{gather*}
z_{k}( 1) =z_{k}( 0) +\varepsilon \dot{z}(0) +O\big( \varepsilon ^{2}\big)
\end{gather*}
%\end{subequations}
(via the second of the three relations (\ref{zepsi}), with $\ell =0$). Via
these three relations (\ref{SolMod1bis}) becomes
%\begin{subequations}
\begin{gather*}
\prod\limits_{k=1}^{N}\left[ \frac{z-\exp (-i\omega t)\big(
1-\varepsilon \omega ^{2}t/2\big) \{ z_{k}( 0)
+\varepsilon [ \dot{z}_{k}( 0) +i\omega z_{k}(
0) ] \} +O\big( \varepsilon ^{2}\big) }{z-\exp
(-i\omega t)\big( 1-\varepsilon \omega ^{2}t/2\big) z_{k} (
0 ) +O\big( \varepsilon ^{2}\big) }\right]  \notag \\
\qquad{} =\frac{\exp (-i\omega t)\big( 1-\varepsilon \omega ^{2}t/2\big)
 ( 1+i\varepsilon \omega  ) -1+O\big( \varepsilon ^{2}\big) }{\exp (-i\omega t)\big( 1-\varepsilon \omega ^{2}t/2\big)
-1+O\left( \varepsilon ^{2}\right) },
\end{gather*}
i.e.\ (dividing each numerator by the corresponding denominator)
\begin{gather*}
\prod\limits_{k=1}^{N}\left[ 1-\varepsilon \frac{[ \dot{z}_{k} (
0 ) +i\omega z_{k}( 0) ] }{z-\exp (-i\omega
t)z_{k}( 0) }+O\big( \varepsilon ^{2}\big) \right]
=1+\varepsilon \frac{i\omega }{\exp (-i\omega t)-1}+O\big(
\varepsilon ^{2}\big) .
\end{gather*}
%\end{subequations}
Clearly to order $\varepsilon ^{0}=1$ this yields the trivial identity $1=1,$
and to order $\varepsilon $ just the formula~(\ref{SolvContAlphaEqOne}). %Q.E.D.

Let us end this subsection  by pointing out that there is another \textit{ansatz} that allows to transform the system of $N^{2}$ equations (\ref{EqMV}) into two separate systems of $N$ equations, but only in the special case $a=1$. This alternative \textit{ansatz} reads (instead of (\ref{Vans}))
\begin{gather*}
Y_{nm}=\frac{f_{m}}{\tilde{z}_{m}-z_{n}},\qquad n,m=1,\dots,N,
\end{gather*}
entailing (but only provided $a=1$) the replacement of the system of $N^{2}$
equations (\ref{EqMV}) with the following two systems of $N$ equations:
%\begin{subequations}
\begin{gather*}
\sum_{k=1}^{N}\left[ \frac{f_{k}}{ ( \tilde{z}_{k}-z_{n} ) ^{2}}\right] =1,\qquad n=1,\dots,N,
\\
\sum_{k=1}^{N}\left[ \frac{f_{k}}{ \big( \widetilde{\tilde{z}}_{n}-\tilde{z}
_{k}\big) ^{2}}\right] =1,\qquad n=1,\dots,N.
\end{gather*}
%\end{subequations}
But, as indicated at the end of  Section~\ref{section1}, we postpone the
treatment of the corresponding class of \textit{discrete-time} dynamical
systems to a separate paper.

\subsection{Solution of the second model}\label{section3.2}

The proof of the f\/indings reported in Subsection~\ref{section2.2} is analogous to that
provided above, see Subsection~\ref{section3.1}, so our treatment in this subsection is
quite terse, being limited to indicate the changes with respect to that
reported in the preceding Subsection~\ref{section3.1}. Now the system of matrix evolution
equations (\ref{EqU}) is generalized to read
\begin{gather}
\tilde{U}=U ( a I+b V ) +V,\qquad \tilde{V}=V;  \label{GenEqU}
\end{gather}%
hence its solution is given by (\ref{GenSolU}). Clearly this evolution
equation, (\ref{GenEqU}), respectively its solution, (\ref{GenSolU}), reduce
to (\ref{EqU}) respectively to (\ref{SolU}) when $b$ vanishes.

The rest of the treatment is analogous. (\ref{EqMZ}) is now generalized to
read
%\begin{subequations}
\begin{gather*}
M\tilde{Z}-aZM=( I+bZ) Y,
\end{gather*}
hence\ it yields, in place of (\ref{MV}),
\begin{gather*}
M_{nm}=\left( \frac{1+bz_{n}}{\tilde{z}_{m}-az_{n}}\right)
Y_{nm},\qquad n,m=1,\dots,N.  %\label{Mmodel2}
\end{gather*}
%\end{subequations}
In place of (\ref{Cond}) (again via the \textit{ansatz} (\ref{Vans})) one
now has
\begin{subequations}
\begin{gather}
\sum_{k=1}^{N}\left[ \frac{g_{k}( 1+b\tilde{z}_{k}) }{\big(
\widetilde{\tilde{z}}_{m}-a\tilde{z}_{k}\big) }-\frac{g_{k} (
1+bz_{n} ) }{\tilde{z}_{k}-az_{n}}\right] =0,\qquad n,m=1,\dots,N,
\end{gather}
hence in place of (\ref{Conda}) and (\ref{Condb}) one gets the two sets of $%
N $ equations%
\begin{gather}
\sum_{k=1}^{N}\left[ \frac{g_{k}}{\tilde{z}_{k}-az_{n}}\right] =\frac{1}{1+bz_{n}},\qquad n=1,\dots,N,  \label{EqgkMod21}
\\
\sum_{k=1}^{N}\left[ \frac{g_{k} ( 1+b\tilde{z}_{k} ) }{\widetilde{\tilde{z}}_{n}-a\tilde{z}_{k}}\right] =1,\qquad n=1,\dots,N.
\label{EqgkMod22}
\end{gather}
\end{subequations}

By solving the f\/irst set one obtains (via  Lemma~\ref{lemmaA2}, with $\xi _{k}=
\tilde{z}_{k},$ $\eta _{n}=az_{n},$ $c_{n}=1/( 1+bz_{n}) $, and
then the identity~(\ref{Identity3}) with $z=-a/b$, $\eta _{k}=az_{k}$, $\zeta _{j}=\tilde{z}_{j}$) the following expression of $g_{k}\equiv \hat{g}_{k}( \underline{z},\underline{\tilde{z}}) $:
\begin{subequations}
\begin{gather}
\hat{g}_{k} ( \underline{z},\underline{\tilde{z}} ) =\left( \frac{
\tilde{z}_{k}-az_{k}}{1+bz_{k}}\right) \prod\limits_{j=1,\; j\neq k}^{N}
\left[ \left( \frac{\tilde{z}_{k}-az_{j}}{\tilde{z}_{k}-\tilde{z}_{j}}
\right) \left( \frac{1+b\tilde{z}_{j}/a}{1+bz_{j}}\right) \right] .
\label{gkMod2Ver1}
\end{gather}

By solving instead the second set one obtains (via  Lemma~\ref{lemmaA1}, with $
\xi _{k}=-a\tilde{z}_{k},$ $\eta _{n}=-\widetilde{\tilde{z}}_{n}$, $c=1$ and
$g_{k}$ replaced by $g_{k}( 1+b\tilde{z}_{k}) $) the following
expression of $g_{k}\equiv \check{g}_{k}\big( \underline{\tilde{z}},
\underline{\widetilde{\tilde{z}}}\big) $:
\begin{gather}
\check{g}_{k}\big( \underline{\tilde{z}},\underline{\widetilde{\tilde{z}}}
\big) =a^{1-N} \left( \frac{\widetilde{\tilde{z}}_{k}-a\tilde{z}_{k}}{1+b
\tilde{z}_{k}}\right) \prod\limits_{j=1,\; j\neq k}^{N}\left( \frac{\widetilde{
\tilde{z}}_{j}-a\tilde{z}_{k}}{\tilde{z}_{j}-\tilde{z}_{k}}\right) .
\label{gkMod2Ver2}
\end{gather}
\end{subequations}

The three versions, (\ref{EqMod2}), of the equations of motion reported in
Subsection~\ref{section2.2} then follow by inserting (\ref{gkMod2Ver1}) in (\ref{EqgkMod22}), by inserting (\ref{gkMod2Ver2}) in (\ref{EqgkMod21}), and by
equating~(\ref{gkMod2Ver1}) to~(\ref{gkMod2Ver2}).

Next, let us prove  Proposition~\ref{proposition2.2.1}. One proceeds again in close
analogy to the treatment of the preceding Subsection~\ref{section3.1}, hence we only
mention where the treatment here dif\/fers from that provided there. It is
easily seen that the expression of~$U ( 0 ) $ is the same as that
given there, see~(\ref{Uzeronm}), while the expression of~$V( 0) $
(because one must now use  Lemma~\ref{lemmaA5} with $f_{n}=1+bz_{n}(0) $ rather than $f_{n}=1$) is now given by~(\ref{A}). The insertion
of these expressions of~$U ( 0 ) $ and~$V ( 0 ) $ in~(\ref{GenEqU}) reproduce~(\ref{SolUMod2}), thereby proving  Proposition~\ref{proposition2.2.1}.

Let us end this subsection by outlining what happens in the \textit{continuous-time} limit which obtains by setting
\begin{gather*}
a=1+\varepsilon b\eta ,\qquad V( 0) =\varepsilon B
\end{gather*}
with $\varepsilon $ inf\/initesimal, and correspondingly replacing the \textit{discrete-time} matrix evolution equation~(\ref{GenEqU}) with the matrix ODE
\begin{subequations}
\begin{gather}
\dot{U}=bU( \eta I+B) +B,  \label{GenODEU}
\end{gather}
the solution of which reads
\begin{gather}
U ( t ) =U ( 0 )  \exp  [ b ( \eta  I+B )  t ] +B [ b ( \eta  I+B )  ] ^{-1}\{ \exp [b( \eta I+B) t] -I\} .
\end{gather}
\end{subequations}
As already mentioned in Subsection~\ref{section2.2}, the (\textit{continuous-time})
goldf\/ish-type model obtainable by focussing appropriately on the evolution
of the $N$ eigenvalues $z_{n}(t)$ of this $N\times N$ matrix~$U( t)$ (evolving according to~(\ref{GenODEU})) was, to the best
of our knowledge,~\textit{new}, when the \textit{solvable} matrix evolution
equation~(\ref{GenODEU}) was identif\/ied as \textit{continuous-time} limit of~(\ref{GenEqU}); its treatment is provided in~\cite{C2011}.

\subsection{Solution of the third model}\label{section3.3}

The starting point is the following linear system of two \textit{discrete-time} matrix evolution equations:
\begin{subequations}
\begin{gather}
\tilde{U}=a_{+}U+\beta V,\qquad \tilde{V}=a_{-}V,  \label{UVtilde}
\end{gather}%
where the $3$ constants $a_{\pm }$, $\beta $ are \textit{a priori} arbitrary ($\beta \neq 0$). It is easily seen that the solution of the initial-value
problem for $U$ reads as follows (with an analogous formula for $V$):
\begin{gather}
U( \ell ) =a_{+}^{\ell }C_{+}+a_{-}^{\ell }C_{-},
\label{SolvU3}
\end{gather}
\end{subequations}
and the two constant matrices $C_{\pm }$ given by~(\ref{C+-1}).

%\textit{Remark 3.3.1}.
\begin{remark}\label{3.3.1}
 This solution $U( \ell ) $ of the
initial-value problem for the \textit{discrete-time} $N\times N$ matrix
evolution equation (\ref{UVtilde}) depends only on the $2$ constants $a_{\pm
}$: see~(\ref{SolvU3}) and~(\ref{C+-1}). Indeed the system of two
f\/irst-order \textit{discrete-time} evolution equations~(\ref{UVtilde}) is
easily seen to correspond to the single second-order evolution equation
\begin{gather*}
\widetilde{\tilde{U}}- ( a_{+}+a_{-} )  \tilde{U}+a_{+}a_{-}U=0,
\end{gather*}
from which the constant $\beta $ has disappeared (but note that this \textit{second-order} matrix ODE obtains only if $\beta \neq 0;$ indeed if $\beta =0$, $U$ satisf\/ies a \textit{first-order} evolution equation, see the f\/irst of
the two equations (\ref{UVtilde}).
\end{remark}

%\textit{Remark 3.3.2}.
\begin{remark}\label{3.3.2}
For $a_{+}=a$, $\beta =1$, $a_{-}=1$, the system (\ref{UVtilde}) coincides with (\ref{EqU}) hence this model reduces to the f\/irst
model, conf\/irming Remark~\ref{remark2.3.1}.
\end{remark}

We then proceed as in the f\/irst part of Section~\ref{section3}. It is then easily seen
that the two matrix evolution equations (\ref{UVtilde}) become
%\begin{subequations}
%\label{EqMYZ1}
\begin{gather*}
M\tilde{Z}=a_{+}ZM+\beta Y,  %\label{MY1}
\\
M\tilde{Y}=a_{-}Y\tilde{M}.
\end{gather*}
%\end{subequations}
Via (\ref{Zdiag}) the f\/irst of these two equations reads, componentwise,
\begin{gather}
M_{nm}=\frac{\beta Y_{nm}}{\tilde{z}_{m}-a_{+}z_{n}},\qquad n,m=1,\dots,N,
\label{MY3}
\end{gather}
and using this formula it is easily seen that the second can be written,
componentwise, as follows:
\begin{gather}
 \sum_{k=1}^{N}\left[ Y_{nk}\tilde{Y}_{km}\left( \frac{a_{-}}{\widetilde{\tilde{z}}_{m}-a_{+}\tilde{z}_{k}}-\frac{1}{\tilde{z}_{k}-a_{+}z_{n}}
\right) \right] =0,  \qquad
 n,m=1,\dots,N.  \label{EqM1}
\end{gather}

At this point we use again the \textit{ansatz} (\ref{Vans}) for the matrix $Y_{nm}$, the consistency of which is vindicated by the subsequent
developments. Here the $N$ quantities $g_{m}$ are again \textit{a priori}
arbitrary; they shall be determined as functions of the $2N$ un-updated and
once-updated coordinates $z_{n}\equiv z_{n}( \ell) $ and $\tilde{z}_{n}\equiv z_{n}( \ell +1) $, or alternatively of the once and
twice updated coordinates $\tilde{z}_{n}\equiv z_{n}( \ell +1) $
and $\widetilde{\tilde{z}}_{n}\equiv z_{n}( \ell +2) $, see
below. It is indeed immediately seen that via the \textit{ansatz}~(\ref{Vans}) the system of~$N^{2}$ equations~(\ref{EqM1}) becomes
\begin{subequations}
\begin{gather}
\sum_{k=1}^{N}\left( \frac{a_{-}g_{k}}{\widetilde{\tilde{z}}_{m}-a_{+}
\tilde{z}_{k}}-\frac{g_{k}}{\tilde{z}_{k}-a_{+}z_{n}}\right)
=0,\qquad n,m=1,\dots,N;
\end{gather}
hence it can be replaced by the following two separated systems of \textit{only }$N$ equations:
\begin{gather}
\sum_{k=1}^{N}\left( \frac{g_{k}}{\tilde{z}_{k}-a_{+}z_{n}}\right)
=1,\qquad n=1,\dots,N,  \label{EqgkMod31}
\\
\sum_{k=1}^{N}\left( \frac{g_{k}}{\widetilde{\tilde{z}}_{n}-a_{+}\tilde{z}
_{k}}\right) =\frac{1}{a_{-}},\qquad n=1,\dots,N.  \label{EqgkMod32}
\end{gather}
\end{subequations}

The f\/irst, (\ref{EqgkMod31}), of these two systems def\/ines uniquely the $N$
quantities $g_{k}\equiv \hat{g}_{k}\left( \underline{z},\underline{\tilde{z}}%
\right) $, yielding again, via (\ref{ff}), the expression (\ref{gk}) (with $%
a $ replaced by $a_{+}$):
\begin{gather}
\hat{g}_{k} ( \underline{z},\underline{\tilde{z}} ) = ( \tilde{z
}_{k}-a_{+}z_{k})  \prod\limits_{j=1,\; j\neq k}^{N}\left( \frac{\tilde{z}_{k}-a_{+}~z_{j}}{\tilde{z}_{k}-\tilde{z}_{j}}\right) ,
\qquad k=1,\dots,N.
\label{gka}
\end{gather}
Insertion of this expression in the second,~(\ref{EqgkMod32}), of the two
systems written just above then yields the evolution equation~(\ref{EvEq1}).
The identif\/ication of the third \textit{discrete-time} dynamical system of
goldf\/ish type, see~(\ref{EvEq1}), is thereby accomplished.

The second version, (\ref{EvEq2a}), of this model obtains by solving for $g_{k}\equiv \check{g}_{k}\big( \underline{\tilde{z}},\underline{\widetilde{\tilde{z}}}\big) $ the se\-cond,~(\ref{EqgkMod32}), of the two systems
written above (via  Lemma~\ref{lemmaA1}, with $\xi _{k}=a_{+}\tilde{z}_{k}$, $
\eta _{n}=\widetilde{\tilde{z}}_{n}$ and $c=-1/a_{-}$), thereby obtaining
\begin{gather}
\check{g}_{k}\big( \underline{\tilde{z}},\underline{\widetilde{\tilde{z}}}
\big) =\left( \frac{\widetilde{\tilde{z}}_{k}-a_{+}\tilde{z}_{k}}{a_{-}}
\right) \prod\limits_{j=1,\; j\neq k}^{N}\left( \frac{a_{+}\tilde{z}_{k}-
\widetilde{\tilde{z}}_{j}}{a_{+}\tilde{z}_{k}-a_{+}\tilde{z}_{j}}\right)
,\qquad k=1,\dots,N;  \label{gkb}
\end{gather}
and by then inserting this expression of $g_{k}$ in (\ref{EqgkMod31}).

And the third version, (\ref{EvEq33}), of this model obtains by equating (\ref{gka}) to~(\ref{gkb}).

As for the proof of  Proposition~\ref{proposition2.3.1}, it follows immediately from
the above treatment, see in particular~(\ref{SolU1}); there only remains to
justify the identif\/ication of the two matrices~$U( 0) $ and~$U( 1) $, see (\ref{Uzero3}) and (\ref{Uone3}).

The f\/irst of the two formulas, (\ref{Uzero3}), is just (\ref{Uzero}).

To prove the second, (\ref{Uone3}), we note that (\ref{Uzero}) and (\ref{Mel}) entail (for $\ell =0$)
\begin{gather}
R( 1) =M( 0) ,  \label{R(1)M(0)}
\end{gather}
while (\ref{UZ}) (with $\ell =0$) reads
\begin{subequations}
\begin{gather}
U ( 1 ) =R ( 1 )  Z ( 1 )  [ R (
1 )  ] ^{-1}.  \label{VY(0)1}
\end{gather}
Hence (via (\ref{R(1)M(0)}))
\begin{gather}
U ( 1 ) =M ( 0 )  Z ( 1 )  [ M (0 )  ] ^{-1}.  \label{Uone}
\end{gather}
\end{subequations}

We now note that (\ref{MY3}) and the \textit{ansatz}~(\ref{Vans}) imply that
the $N\times N$ matrix $M( 0) $ is def\/ined componentwise as
follows:
\begin{subequations}
\begin{gather}
M_{nm} ( 0 ) =\frac{g_{m} ( 0 ) }{z_{m} ( 1 )
-a_{+} z_{n} ( 0 ) },\qquad n,m=1,\dots,N,  \label{Mnm(0)1}
\end{gather}
with (see (\ref{gka}))
\begin{gather}
g_{m} ( 0 ) = [ z_{m} ( 1 ) -a_{+} z_{m} (
0 )  ]  \prod\limits_{j=1,\; j\neq m}^{N}\left[ \frac{z_{m} (
1 ) -a_{+}z_{j} ( 0 ) }{z_{m} ( 1 ) -z_{j} (
1 ) }\right] ,\qquad m=1,\dots,N,  \label{gm3}
\end{gather}
so that
\begin{gather}
M_{nm}( 0) =\frac{z_{m}( 1) -a_{+}z_{m}(
0) }{z_{m}( 1) -a_{+}z_{n}( 0) }
 \prod\limits_{j=1,\; j\neq m}^{N}\left[ \frac{z_{m} ( 1 )
-a_{+} z_{j} ( 0 ) }{z_{m} ( 1 ) -z_{j} ( 1 ) }
\right] ,\qquad n,m=1,\dots,N.  \label{Mnm3}
\end{gather}

Next, we note that the expression (\ref{Mnm(0)1}) of the matrix $M(0) $ entails~-- via  Lemma~\ref{lemmaA5} (with $f_{n}=1$, $g_{m}=g_{m}( 0) $, $\xi _{m}=z_{m}( 1) $, $\eta
_{n}=a_{+}z_{n}( 0) $) and the expression of $g_{m}(0) $ given above~-- that its inverse, appearing in the
right-hand-side of~(\ref{Uone}), is explicitly given, componentwise, as
follows:
\begin{gather}
 \big\{  [ M ( 0 )  ] ^{-1}\big\} _{nm}  = a_{+}^{1-N}
\left[ \frac{a_{+} z_{m} ( 0 ) -z_{m} ( 1 ) }{a_{+} z_{m} ( 0 ) -z_{n} ( 1 ) }\right] \prod
\limits_{j=1,\;j\neq m}^{N}\left[ \frac{a_{+} z_{m}( 0)
-z_{j}( 1) }{z_{m}( 0) -z_{j}( 0) }\right]
,  \notag \\
n,m =1,\dots,N.
\end{gather}
\end{subequations}
The insertion of this formula and (\ref{Mnm3}) in (\ref{Uone}) entails that
the matrix $U(1) $ reads componentwise as follows:
\begin{gather*}
U_{nm} ( 1 ) =a_{+}^{1-N} \prod\limits_{j=1,\; j\neq m}^{N}\left[
\frac{a_{+}z_{m} ( 0 ) -z_{j} ( 1 ) }{z_{m} ( 0 )
-z_{j} ( 0 ) }\right]    \notag \\
\phantom{U_{nm} ( 1 ) =}{}
\times \sum_{k=1}^{N}\left\{ z_{k}( 1) \frac{\prod%
\limits_{j=1,\; j\neq n}[ z_{k}( 1) -a_{+}z_{j}(
0) ] }{\prod\limits_{j=1,\; j\neq k}[ z_{k}( 1)
-z_{j}( 1)] }\right\} ,\qquad n,m =1,\dots,N.
\end{gather*}
And it is then immediately seen that this formula yields, via the identity~(\ref{Identity5}) (with $\eta _{k}=z_{k} (1) $, $k=1,\dots,N$; $\zeta _{j}=a_{+}z_{j} ( 0) $, $j=1,\dots,n-1,n+1,\dots,N$), the
formula~(\ref{Uone3}). % Q. E. D.

\subsection{Solution of the fourth model}\label{section3.4}

The treatment in this subsection  is rather terse, since it is analogous
to that of the preceding subsections; and the notation is of course
analogous. But now the starting point is the following nonlinear system of
two \textit{discrete-time} matrix evolution equations:
\begin{gather}
\tilde{U}=\alpha U+\beta V+\eta UV,\qquad \tilde{V}=\rho +\gamma V,
\label{EqUV4}
\end{gather}
featuring the 5 arbitrary constants $\alpha $, $\beta $, $\eta $, $\rho $, $\gamma $ (which, as noted in Subsection~\ref{section2.4}, can be reduced to 4 by taking
advantage of the freedom to rescale~$V$).

It is a standard task to see that the solution of the initial-value problem
for this matrix system reads as follows:
\begin{gather*}
V( \ell) =\gamma ^{\ell }V( 0) +\rho \frac{\gamma
^{\ell }-1}{\gamma -1}I,  %\label{Vel2}
\end{gather*}
with $U( \ell) $ given by (\ref{Usol4}).

We now proceed again as in the f\/irst part of Section~\ref{section3}. Via~(\ref{Mel}) and~(\ref{VMR}) we get from~(\ref{EqUV4}) the two matrix equations
\begin{subequations}
\begin{gather}
M \tilde{Z}=\alpha ZM+( \beta +\eta Z) Y,  \label{MZ2}
\\
M\tilde{Y}=\rho M\tilde{M}+\gamma Y\tilde{M}.  \label{MY2}
\end{gather}
\end{subequations}
The f\/irst, (\ref{MZ2}), of these two matrix equations entails,
componentwise,
\begin{gather*}
Y_{nm}=\left( \frac{\tilde{z}_{m}-\alpha z_{n}}{\eta z_{n}+\beta }\right)
M_{nm},\qquad n,m=1,\dots,N;
\end{gather*}
hence the second, (\ref{MY2}), of these two matrix equations yields (when
written componentwise) the following $N^{2}$ equations:
\begin{subequations}
\begin{gather}
\sum_{k=1}^{N}\left[ M_{nk}\tilde{M}_{km}\left( \frac{\widetilde{\tilde{z}}_{m}-\alpha \tilde{z}_{k}}{\eta \tilde{z}_{k}+\beta }-\rho -\gamma
\frac{\tilde{z}_{k}-\alpha z_{n}}{\eta z_{n}+\beta }\right) \right] =0,
\qquad n,m=1,\dots,N,
\end{gather}
which, as can be easily verif\/ied, can be conveniently rewritten as follows:
\begin{gather}
\sum_{k=1}^{N}\left[ M_{nk}\tilde{M}_{km}\left( \frac{\widetilde{\tilde{z}}_{m}-a\tilde{z}_{k}-b}{\eta \tilde{z}_{k}+\beta }-\gamma \frac{\tilde{z}_{k}-az_{n}-b}{\eta z_{n}+\beta }\right) \right] =0,  \qquad n,m=1,\dots,N,  \label{EqM2}
\end{gather}
\end{subequations}
with the two constants $a$ and $b$ def\/ined by (\ref{ab}).

Next, we make the following \textit{ansatz} for the matrix $M_{nm}$:
\begin{gather}
M_{nm}=\frac{g_{m}}{\tilde{z}_{m}-az_{n}-b},\qquad n,m=1,\dots,N.  \label{AnsM2}
\end{gather}%
Here the $N$ quantities $g_{m}$ are \textit{a priori} arbitrary; they shall
be determined as functions of the $2N$ un-updated and once-updated
coordinates $z_{n}\equiv z_{n}( \ell) $ and $\tilde{z}_{n}\equiv
z_{n}( \ell +1) $, or of the once and twice updated coordinates $
\tilde{z}_{n}\equiv z_{n}( \ell +1) $ and $\widetilde{\tilde{z}}
_{n}\equiv z_{n}( \ell +2) $, see below. It is indeed immediately
seen that via this \textit{ansatz}~(\ref{AnsM2}) the system of~$N^{2}$
equations~(\ref{EqM2}) can be rewritten as follows:
\begin{subequations}
\begin{gather}
 \sum_{k=1}^{N}\left[ \frac{g_{k}( \eta z_{n}+\beta ) }{(
\eta \tilde{z}_{k}+\beta ) ( \tilde{z}_{k}-az_{n}-b) }
-\gamma \frac{g_{k}}{\big( \widetilde{\tilde{z}}_{m}-a \tilde{z}_{k}-b\big) }\right] =0,  \qquad
 n,m=1,\dots,N;
\end{gather}
hence it can be replaced by the following two separated systems of \textit{%
only }$N$ equations:
\begin{gather}
\sum_{k=1}^{N}\left[ \frac{g_{k}}{ ( \eta  \tilde{z}_{k}+\beta  )
 ( \tilde{z}_{k}-az_{n}-b ) }\right] =\frac{1}{ ( \eta
 z_{n}+\beta  ) },\qquad n=1,\dots,N,  \label{Eqg2}
\\
\sum_{k=1}^{N}\left( \frac{g_{k}}{\widetilde{\tilde{z}}_{n}-a \tilde{z}_{k}-b}\right) =\frac{1}{\gamma },\qquad n=1,\dots,N.
\end{gather}
\end{subequations}
The f\/irst of these two systems def\/ines uniquely the~$N$ quantities $g_{k}\equiv \hat{g}_{k} ( \underline{z},\underline{\tilde{z}} ) $,
yielding their expression~(\ref{gnhat4}) (see Appendix~\ref{appendixB} for a proof). The
second equation then yields the evolution equation~(\ref{EvEq2}).

The alternative possibility is to determine (via  Lemma~\ref{lemmaA1}, with $%
\xi _{k}=a~\tilde{z}_{k},~\eta _{n}=\widetilde{\tilde{z}}_{n}-b,~c=-1/\gamma
$) the quantities $g_{k}\equiv \check{g}_{k}\big( \underline{\tilde{z}},
\underline{\widetilde{\tilde{z}}}\big) $ as solutions of the second
system, yielding the formula~(\ref{gkMod42}); and to then insert these
expressions of $g_{k}\equiv \check{g}_{k}\big( \underline{\tilde{z}},%
\underline{\widetilde{\tilde{z}}}\big) $ in the f\/irst system of equations.
Clearly in this manner one arrives at the equations of motion (\ref{EqMotion4bis}).

And a third possibility is of course to equate (\ref{gnhat4}) to (\ref{gkMod42}), see (\ref{Eqgg}).

The identif\/ication of the three variants of the equations of motion of the
fourth \textit{discrete-time} dynamical system of goldf\/ish type, see
Subsection~\ref{section2.4}, is thereby accomplished.

The proof of  Proposition~\ref{proposition2.4.1} follows immediately from the above
treatment; and we trust that the identif\/ication in terms of the $2N$ initial
data $z_{n}( 0) $ and $z_{n}( 1) $ of the two matrices~$U( 0) $ and~$V( 0) $, see~(\ref{U(0)2}) and~(\ref{V(0)2}), is suf\/f\/iciently obvious (also in the light of the analogous
treatment in the preceding subsections of this section) not to require an
explicit justif\/ication here.

We end this subsection with a terse mention of the \textit{continuous-time}
model that obtains from that treated above (in this subsection) via the
limiting transition from \textit{discrete} to \textit{continuous} time. The
point of departure for the treatment of the \textit{continuous-time}
dynamical system of goldf\/ish type obtained in this manner is the following
\textit{continuous-time} system of two f\/irst-order matrix evolution
equations
\begin{gather*}
\dot{U}=a_{1}U+a_{2}V+a_{3}UV,\qquad \dot{V}=a_{4}+a_{5}V,
\end{gather*}
that obtains from (\ref{EqUV4}) via the assignments $t\Rightarrow
\varepsilon  \ell $, $U( \ell ) \Rightarrow U( t) $, $V ( \ell  ) \Rightarrow V ( t ) $, $\alpha =1+\varepsilon
a_{1}$, $\beta =\varepsilon a_{2}$, $\eta =\varepsilon a_{3}$, $\rho
=\varepsilon a_{4}$, $\gamma =1+\varepsilon a_{5}$, with $\varepsilon $
inf\/initesimal. The resulting model of goldf\/ish type was, to the best of our
knowledge, \textit{new}; a detailed treatment of it is provided in~\cite{C2011c}.

\section{Outlook}\label{section4}

In this paper we have introduced and tersely analyzed 4 dif\/ferent \textit{discrete-time} dynamical systems of goldf\/ish type. The possibility to
identify other \textit{discrete-time} evolution equations amenable to exact
treatment by variations of the methodology used in this paper is open: let
us outline here an avenue to such generalizations.

Consider the system of two $N\times N$ matrix \textit{discrete-time}
f\/irst-order evolution equations
\begin{subequations}
\label{SysOut}
\begin{gather}
\sum_{s=1}^{S_{1}} \big[ F_{1,s} ( U )  F_{2,s} ( \tilde{U}
 )  \big] =\sum_{s=1}^{S_{2}}\big[ \Phi _{1,s} ( U ) V\Phi
_{2,s}\big( \tilde{U}\big) \big] ,  \label{EqFF}
\\
\sum_{s=1}^{S_{3}}\big[ F_{3,s} ( U )  F_{4,s}\big( \tilde{U}\big) \big] \tilde{V}=\Phi _{3} ( U )  V \Phi _{4}\big(
\tilde{U}\big) ,  \label{EqVV}
\end{gather}
\end{subequations}
where the two $N\times N$ matrices $U\equiv U ( \ell  ) $ and $
V ( \ell  ) $ are the dependent variables, $\ell =0,1,2,\dots$ is the
independent \textit{discrete-time} variable, $S_{1},$ $S_{2},$ $S_{3}$ are 3
arbitrary positive integers, $F_{1,s} ( u ) $, $F_{2,s} (
u ) $, $F_{3,s} ( u ) $, $F_{4,s} ( u ) $ and $\Phi
_{1,s} ( u ) $, $\Phi _{2,s} ( u ) $, $\Phi _{3} (
u ) $, $\Phi _{4} ( u ) $ are $2 (
S_{1}+S_{2}+S_{3}+1 ) $ \textit{a priori} arbitrary (scalar) functions
of their (scalar) argument $u$ (of course becoming $N\times N$ matrices when
the scalar $u$ is replaced by an $N\times N$ matrix). Then introduce the
eigenvalues $z_{n} ( \ell  ) $ of the matrix $U ( \ell  ) $, as well as the matrices $Z\equiv Z ( \ell  ) $, $R\equiv R (
\ell  ) $, $M\equiv M ( \ell  ) $ and $Y\equiv Y ( \ell
 ) $, as above (see (\ref{UZ}) and (\ref{MVel})). It is then plain that
the matrix evolution equation (\ref{EqFF}) becomes
\begin{subequations}
\begin{gather}
\sum_{s=1}^{S_{1}}\big[ F_{1,s} ( Z ) M F_{2,s}\big( \tilde{Z}
\big) \big] =\sum_{s=1}^{S_{2}}\big[ \Phi _{1,s} ( Z ) Y\Phi
_{2,s}\big( \tilde{Z}\big) \big]
\end{gather}
entailing componentwise
\begin{gather}
M_{nm}=\frac{\sum\limits_{s=1}^{S_{2}}\big[ \Phi _{1,s} ( z_{n} )  \Phi
_{2,s} ( \tilde{z}_{m} ) \big] }{\sum\limits_{s=1}^{S_{1}}\big[
F_{1,s} ( z_{n} ) F_{2,s}\big( \tilde{z}_{m}\big) \big] }
Y_{nm},\qquad n,m=1,\dots,N.  \label{MYnm}
\end{gather}
\end{subequations}
Likewise (\ref{EqVV}) becomes
%\begin{subequations}
\begin{gather*}
\sum_{s=1}^{S_{3}}\big[ F_{3,s} ( Z ) M F_{4,s} (
\tilde{Z} ) \big] \tilde{Y}=\Phi _{3} ( Z )  Y \Phi
_{4}\big( \tilde{Z}\big)  \tilde{M},
\end{gather*}
entailing componentwise (via (\ref{MYnm}))
\begin{gather*}
 \sum_{k=1}^{N}\left\{ Y_{nk} \tilde{Y}_{km} \left[
\sum_{s=1}^{S_{2}}F_{3,s} ( z_{n} )  F_{4,s} ( \tilde{z}_{k} ) \right] \frac{\sum\limits_{\sigma =1}^{S_{2}}\big[ \Phi _{1,\sigma
} ( z_{n} ) \Phi _{2,\sigma } ( \tilde{z}_{k} ) \big]}{\sum\limits_{\sigma =1}^{S_{1}}\big[ F_{1,\sigma } ( z_{n} )  F_{2,\sigma
} ( \tilde{z}_{k} ) \big] }\right\}  \notag \\
\qquad{} = \Phi _{3} ( z_{n} )  \sum_{k=1}^{N}\left\{ Y_{nk} \tilde{Y}
_{km} \Phi _{4} ( \tilde{z}_{k} )  \frac{\sum\limits_{s=1}^{S_{2}}\big[
\Phi _{1,s} ( \tilde{z}_{k} )  \Phi _{2,s}\big( \widetilde{\tilde{z}}_{m}\big) \big] }{\sum\limits_{s=1}^{S_{1}}\big[ F_{1,s} ( \tilde{z}_{k} ) F_{2,s}\big( \widetilde{\tilde{z}}_{m}\big) \big] }
 \right\} ,  \qquad
n,m  = 1,\dots,N.
\end{gather*}
%\end{subequations}

And via the \textit{ansatz} $Y_{nm}=g_{m}$ (see (\ref{Vans})) this system of
$N^{2}$ equations can clearly be replaced by the following two systems of $N$
\textit{linear }algebraic equations for the $N$ quantities $g_{k}$:
\begin{subequations}
\label{Eqgk}
\begin{gather}
 \sum_{k=1}^{N}\left\{ g_{k}\left[ \sum_{s=1}^{S_{2}}F_{3,s} (
z_{n} )  F_{4,s} ( \tilde{z}_{k} ) \right] \frac{\sum\limits_{\sigma
=1}^{S_{2}}\big[ \Phi _{1,\sigma } ( z_{n} )  \Phi _{2,\sigma
} ( \tilde{z}_{k} ) \big] }{\sum\limits_{\sigma =1}^{S_{1}}\big[
F_{1,\sigma } ( z_{n} )  F_{2,\sigma } ( \tilde{z}_{k} )
\big] }\right\} =\Phi _{3} ( z_{n} ) ,  \qquad\!
n =1,\dots,N,  \!\!\!\!\!\label{Eqgka}
\\
\sum_{k=1}^{N}\left\{ g_{k}\Phi _{4} ( \tilde{z}_{k} )  \frac{\sum\limits_{s=1}^{S_{2}}\big[ \Phi _{1,s} ( \tilde{z}_{k} ) \Phi
_{2,s}\big( \widetilde{\tilde{z}}_{n}\big) \big] }{\sum\limits_{s=1}^{S_{1}}
\big[ F_{1,s} ( \tilde{z}_{k} )  F_{2,s}\big( \widetilde{\tilde{z}}_{n}\big) \big] }\right\} =1,\qquad n=1,\dots,N.  \label{Eqgkb}
\end{gather}
\end{subequations}

One can then solve the f\/irst, (\ref{Eqgka}), respectively the second,~(\ref{Eqgkb}), of these two \textit{linear} systems for the $N$ quantities $%
g_{k}, $ getting the expressions $g_{k}=\hat{g}_{k} ( \underline{z},\underline{\tilde{z}} ) $ respectively $g_{k}=\check{g}_{k}\big(
\underline{\tilde{z}},\widetilde{\underline{\tilde{z}}}\big)$. One
arrives thereby at three \textit{equivalent} systems of $N$ \textit{discrete-time} second-order evolution equations for the $N$ coordinates $z_{n}\left( \ell \right) $: $(i)$ by inserting the expression $\hat{g}_{k} ( \underline{z},\underline{\tilde{z}} )$ in~(\ref{Eqgkb});
$(ii)$ by inserting the expression $\check{g}_{k}\big( \underline{\tilde{z}},\widetilde{\underline{\tilde{z}}}\big) $ in~(\ref{Eqgka}); $(iii)$ by
setting $\hat{g}_{n} ( \underline{z},\underline{\tilde{z}} ) =
\check{g}_{n}\big( \underline{\tilde{z}},\widetilde{\underline{\tilde{z}}}\big) $. While of course the evolution in the discrete time $\ell $
entailed by these equations of motions corresponds to the evolution of the
eigenvalues of the matrix $U ( \ell  ) $ solution of the matrix
evolution equation~(\ref{EqFF}) (with the \textit{ansatz}~(\ref{Vans})
properly taken into account). Hence if that matrix evolution equation, (\ref{EqFF}), is \textit{solvable}, the \textit{discrete-time} dynamical system
described by these three equivalent sets of second-order evolution equations
is as well \textit{solvable}.

One has thereby identif\/ied a \textit{discrete-time} solvable dynamical
system. A remaining open question is the extent to which its equations of
motion can be exhibited in reasonably neat form: this depends on the extent
that the two quantities $g_{k}\equiv \hat{g}_{k}( \underline{z},
\underline{\tilde{z}} )$ respectively $g_{k}\equiv \check{g}_{k}\big(
\underline{\tilde{z}},\widetilde{\underline{\tilde{z}}}\big) $ def\/ined as
solutions of the two \textit{linear} systems (\ref{Eqgka}) respectively (\ref{Eqgkb}) can be expressed more explicitly than via the standard Cramer
formula (ratio of two determinants).

Clearly the 4 models treated in this paper belong to this class (\ref{SysOut}) (the last one, however, only if $\rho =0$): see~(\ref{EqU}), (\ref{GenEqU}), (\ref{UVtilde}) and (\ref{EqUV4}). The interest of additional models
treatable via this approach depends on the neatness of the corresponding
equations of motion, which can only be investigated on a case-by-case basis.

\appendix

\section{Appendix}\label{appendixA}

In this appendix we collect various mathematical developments whose
treatment in the body of the paper would interrupt the f\/low of the
presentation.

First of all we report several mathematical identities. We consider all of
them well-known, but for completeness we either prove them below, or
indicate where proofs can be found. These formulas feature sets of~$N$
numbers such as~$\xi _{n}$ or~$\eta _{n}$ or~$\zeta _{n}$; these numbers are
\textit{arbitrary} but for simplicity we assume them to be \textit{distinct}. The formulas of course remain valid when these numbers are not distinct,
but possibly only by taking appropriate limits. Sometimes an \textit{arbitrary} number $z$ also appears.
%\begin{subequations}
\begin{gather}
\sum_{k=1}^{N}\left[ \prod\limits_{j=1,\; j\neq k}^{N}\left( \frac{\zeta _{j}-z
}{\zeta _{j}-\zeta _{k}}\right) \right] =1,  \label{Identity1}
\\
\sum_{k=1}^{N}\left[ \frac{1}{\zeta _{k}} \prod\limits_{j=1,\; j\neq
k}^{N}\left( \frac{1}{\zeta _{j}-\zeta _{k}}\right) \right]
=\prod\limits_{j=1}^{N}\left( \frac{1}{\zeta _{j}}\right) ,
\label{Identity1a}
\\
%\end{gather}
%\end{subequations}
%\begin{subequations}
%\begin{gather}
\sum_{k=1}^{N}\left[ \zeta _{k}^{n-1} \prod\limits_{j=1,\; j\neq k}^{N}\left(
\frac{\zeta _{j}-z}{\zeta _{j}-\zeta _{k}}\right) \right]
=z^{n-1},\qquad n=1,2,\dots,N,  \label{Identity1b}
\\
\sum_{k=1}^{N}\left[ \zeta _{k}^{n-1} \prod\limits_{j=1,j\neq k}^{N}\left(
\frac{1}{\zeta _{k}-\zeta _{j}}\right) \right] =\delta
_{nN},\qquad n=1,2,\dots,N,  \label{Identity1c}
\\
%\end{gather}
%\end{subequations}
%\begin{gather}
\prod\limits_{j=1,\; j\neq m}^{N}\left( \frac{\zeta _{j}-\zeta _{n}}{\zeta
_{j}-\zeta _{m}}\right) =\delta _{nm},\qquad n,m=1,2,\dots,N,  \label{Identity2}
\\
%\end{gather}%
%\begin{subequations}
%\begin{gather}
 \sum_{k=1}^{N}\left\{ \left[ \prod\limits_{j=1,\; j\neq k}^{N}\!\!\left( \frac{
\xi _{j}-\eta _{n}}{\xi _{j}-\xi _{k}}\right) \right]  \left[
\prod\limits_{j=1,\; j\neq m}^{N}\!\!\left( \frac{\eta _{j}-\xi _{k}}{\eta
_{j}-\eta _{m}}\right) \right] \right\}  =\delta _{nm},\!\qquad n,m=1,2,\dots,N,\!\!\!  \label{IdentityQI}
\\
 \sum_{k=1}^{N}\left[ \left( \frac{1}{z-\eta _{k}}\right) \frac{\prod\limits_{j=1,\; j\neq n}^{N}\left( \zeta _{j}-\eta _{k}\right) }{\prod\limits_{j=1,\; j\neq k}^{N}\left( \eta _{j}-\eta _{k}\right) }\right]
=\frac{\prod\limits_{j=1,\; j\neq n}^{N}( \zeta _{j}-z) }{\prod\limits_{j=1}^{N}( \eta _{j}-z) }\equiv \frac{1}{z-\zeta
_{n}}\prod\limits_{j=1}^{N}\left( \frac{\zeta _{j}-z}{\eta _{j}-z}\right) ,
\notag \\
n =1,2,\dots,N,  \label{Identity3}\\
%\end{gather}
%\end{subequations}
%\begin{subequations}
%\begin{gather}
 \sum_{k=1}^{N}\left[ \frac{\prod\limits_{j=1,\; j\neq n}^{N}\left( \zeta
_{j}-\eta _{k}\right) }{\prod\limits_{j=1,\; j\neq k}^{N}\left( \eta
_{j}-\eta _{k}\right) }\right] \equiv \sum_{k=1}^{N}\left[ \left( \frac{\eta
_{k}-\zeta _{k}}{\eta _{k}-\zeta _{n}}\right) \prod\limits_{j=1,\; j\neq
k}^{N}\left( \frac{\zeta _{j}-\eta _{k}}{\eta _{j}-\eta _{k}}\right) \right]
=1, \notag\\
 n=1,2,\dots,N,  \label{Identity3a}
\\
 \sum_{k=1}^{N}\left[ \left( \frac{1}{\eta _{k}-\xi _{n}}\right)
 \prod\limits_{j=1,\; j\neq k}^{N}\left( \frac{1}{\eta _{j}-\eta _{k}}\right)
\right] =\prod\limits_{j=1}^{N}\left( \frac{1}{\eta _{j}-\xi _{n}}\right) ,
\qquad n =1,2,\dots,N,  \label{Identity3d}
\\
 \sum_{k=1}^{N}\left[ \left( \frac{\eta _{k}-\zeta _{k}}{\eta _{k}-z}
\right)  \prod\limits_{j=1,\; j\neq k}^{N}\left( \frac{\zeta _{j}-\eta _{k}}{\eta _{j}-\eta _{k}}\right) \right]
\equiv \sum_{k=1}^{N}\left[ \frac{\prod\limits_{j=1}^{N}\left( \zeta
_{j}-\eta _{k}\right) }{\left( z-\eta _{k}\right) \prod\limits_{j=1,\; j\neq
k}^{N}\left( \eta _{j}-\eta _{k}\right) }\right]  \notag \\
\qquad  = 1-\prod\limits_{j=1}^{N}\left( \frac{\zeta _{j}-z}{\eta _{j}-z}\right) ,
\label{Identity3b}\\
%\end{gather}
%\end{subequations}
%\begin{gather}
\sum_{k=1}^{N}\left[ \frac{\prod\limits_{j=1}^{N}\left( \zeta _{j}-\eta
_{k}\right) }{\prod\limits_{j=1,\; j\neq k}^{N}\left( \eta _{j}-\eta
_{k}\right) }\right] =\sum_{k=1}^{N}\left( \zeta _{k}-\eta _{k}\right) ,
\label{Identity4}
\\
\sum_{k=1}^{N}\left[ \frac{\eta _{k} \prod\limits_{j=1,\; j\neq n}^{N}\left(
\zeta _{j}-\eta _{k}\right) }{\prod\limits_{j=1,\; j\neq k}^{N}\left( \eta
_{j}-\eta _{k}\right) }\right] =\sum_{k=1,\; k\neq n}^{N}\left( \zeta
_{k}\right) -\sum_{k=1}^{N}\left( \eta _{k}\right) ,\qquad n=1,\dots,N.
\label{Identity5}
\end{gather}

The identity (\ref{Identity1}) (with $z$ an arbitrary number) is implied by
the fact that its left-hand side is a polynomial in $z$ of degree less than $%
N$ (in fact, of degree at most $N-1$) which clearly has the value \textit{%
unity} at the $N$ points $\zeta _{n}$, and the right-hand side, i.e.\
\textit{unity}, is the \textit{unique} polynomial of degree less than $N$ in
$z$ that has the value \textit{unity} in $N$ distinct points. The identity (\ref{Identity1a}) is the special case of (\ref{Identity1}) with $z=0$. The
identity (\ref{Identity1b}) coincides with equation~(2.4.2-32) of \cite{C2001}
(or, as above, it is implied by the observation that its left-hand side is a
polynomial in~$z$ of degree less than $N$ the values of which at the $N$
points $\zeta _{k}$ coincide with the values of the polynomial $z^{n}$ at $z=\zeta _{k}$). The identity (\ref{Identity1c}) coincides with equations~(2.4.3-12) and~\mbox{(2.4.3-21)} of~\cite{C2001}. The identity (\ref{Identity2}) is
obvious. The identities (\ref{IdentityQI}) respectively (\ref{Identity3})
coincide with equations~(2.4.2-26) respectively~(2.4.2-27) of~\cite{C2001} (via
the def\/inition~(2.4.2-24), with $x_{n}=\xi _{n},$ $y_{n}=\eta _{n},$
respectively $x_{n}=z$, $y_{n}=\eta _{n}$, $z_{n}=\zeta _{n}$). The
identities (\ref{Identity3a}) respectively~(\ref{Identity3d}) follow from (\ref{Identity3}) in the limit $z\rightarrow \infty $ respectively $\zeta
_{j}\rightarrow \infty $. The identity~(\ref{Identity3b}) follows from~(\ref{Identity3a}) and~(\ref{Identity3}) via the trivial identity
\begin{gather*}
\frac{\zeta _{n}-\eta _{k}}{z-\eta _{k}}\equiv 1-\frac{\zeta _{n}-z}{\eta_{k}-z}.
\end{gather*}%
Finally, the identity~(\ref{Identity4}) follows from~(\ref{Identity3b}) in
the limit $z\rightarrow \infty ,$ and the identity~(\ref{Identity5}) is just
the special case of the preceding identity~(\ref{Identity4}) with $\zeta
_{n}=0$.

Next we report a simple lemma (for a neat proof see for instance~\cite{S2005}, or below, after the proof of  the following  Lemma~\ref{lemmaA2}).

%\textit{Lemma A1}.
\begin{lemma}\label{lemmaA1}
The solution of the set of $N$ linear algebraic equations
for the $N$ variables $g_{k}$ reading{\samepage
\begin{subequations}
\label{ff}
\begin{gather}
\sum_{k=1}^{N}\frac{g_{k}}{\xi _{k}-\eta _{n}}=c,\qquad n=1,\dots,N
\end{gather}%
is provided by the formula
\begin{gather}
g_{k}=c \left( \xi _{k}-\eta _{k}\right) \prod\limits_{j=1,\; j\neq
k}^{N}\left( \frac{\xi _{k}-\eta _{j}}{\xi _{k}-\xi _{j}}\right)
,\qquad k=1,\dots,N.  \label{gkLemma1}
\end{gather}
\end{subequations}}
\end{lemma}

A generalization of this lemma reads as follows:

%\textit{Lemma A2}.
\begin{lemma}\label{lemmaA2}
The solution of the set of $N$ linear algebraic equations
for the $N$ variables $g_{k}$ reading
\begin{subequations}
\label{fff}
\begin{gather}
\sum_{k=1}^{N}\frac{g_{k}}{\xi _{k}-\eta _{n}}=c_{n},\qquad n=1,\dots,N
\label{Solvf}
\end{gather}
is provided by the formula%
\begin{gather}
g_{k} =\sum_{s=1}^{N}\left\{ c_{s} ( \xi _{k}-\eta _{s} )  \left[
\prod\limits_{j=1,\; j\neq k}^{N}\left( \frac{\xi _{j}-\eta _{s}}{\xi _{j}-\xi
_{k}}\right) \right]  \left[ \prod\limits_{j=1,\; j\neq s}^{N}\left( \frac{
\eta _{j}-\xi _{k}}{\eta _{j}-\eta _{s}}\right) \right] \right\} ,  \notag
\\
n  = 1,\dots,N,  \label{Solvff}
\end{gather}
or equivalently
\begin{gather}
g_{k}  =  ( \xi _{k}-\eta _{k} )  \left[ \prod\limits_{j=1,\; j\neq
k}^{N}\left( \frac{\eta _{j}-\xi _{k}}{\xi _{j}-\xi _{k}}\right) \right]
 \sum_{s=1}^{N}\left\{ c_{s} \left[ \frac{\prod\limits_{j=1,\; j\neq
k}^{N} ( \xi _{j}-\eta _{s} ) }{\prod\limits_{j=1,\; j\neq
s}^{N} ( \eta _{j}-\eta _{s} ) }\right] \right\} ,  \notag \\
n  = 1,\dots,N.  \label{Solvfff}
\end{gather}
\end{subequations}
\end{lemma}

To prove this formula one inserts this expression, (\ref{Solvff}), of $g_{k}$
in (\ref{Solvf}), and notes that one obtains thereby an equality provided
there holds the formula
%\begin{subequations}
\begin{gather*}
\sum_{k=1}^{N}\left\{ \frac{\xi _{k}-\eta _{s}}{\xi _{k}-\eta _{n}} \left[
\prod\limits_{j=1,\; j\neq k}^{N}\left( \frac{\xi _{j}-\eta _{s}}{\xi _{j}-\xi
_{k}}\right) \right]  \left[ \prod\limits_{j=1,\; j\neq s}^{N}\left( \frac{\eta _{j}-\xi _{k}}{\eta _{j}-\eta _{s}}\right) \right] \right\} =\delta
_{sn}
\end{gather*}
or, equivalently,
\begin{gather*}
\sum_{k=1}^{N}\left\{ \left[ \prod\limits_{j=1,\; j\neq k}^{N}\left( \frac{\xi
_{j}-\eta _{n}}{\xi _{j}-\xi _{k}}\right) \right]  \left[ \prod\limits_{j=1,\; j\neq s}^{N}\left( \frac{\eta _{j}-\xi _{k}}{\eta _{j}-\eta
_{s}}\right) \right] \right\} =\delta _{sn} \prod\limits_{j=1}^{N}\left(
\frac{\xi _{j}-\eta _{n}}{\xi _{j}-\eta _{s}}\right) .
\end{gather*}
%\end{subequations}
Clearly this formula is implied by the identity (\ref{IdentityQI}). Lemma~\ref{lemmaA2} is thus proven.

Note that, by setting $c_{n}=c$ in (\ref{Solvfff}) and using the identity (\ref{Identity3a}) (with the dummy index~$k$ replaced by~$s$ and the index~$n$
replaced by~$k$), one reobtains (\ref{gkLemma1}), thereby proving  Lemma~\ref{lemmaA1}.

We now report, and prove, 3 other lemmata.

%\textit{Lemma A3}.
\begin{lemma}\label{lemmaA3}
There holds the formula
\begin{subequations}
\begin{gather}
\det  [ I+X ] =1+\sum_{k=1}^{N}x_{k},  \label{DetDet}
\end{gather}
provided $I$ is the $N\times N$ unit matrix and the $N\times N$ matrix $X$
is defined componentwise as follows:
\begin{gather}
X_{nm}=x_{m},\qquad n,m=1,\dots,N.  \label{XnmLemmaA3}
\end{gather}
\end{subequations}
\end{lemma}

This (presumably well-known) formula is easily proven by recursion.

%\textit{Lemma A4}.
\begin{lemma}\label{lemmaA4}
The $N$ eigenvalues of the $N\times N$ matrix
\begin{subequations}
\begin{gather}
U_{nm}=\delta _{nm}\zeta _{n}+\eta _{m},\qquad  n,m=1,\dots,N,  \label{UUU}
\end{gather}%
coincide with the $N$ solutions of the following algebraic equation in $z$:
\begin{gather}
\sum_{k=1}^{N}\left( \frac{\eta _{k}}{z-\zeta _{k}}\right) -1=0,
\end{gather}
\end{subequations}
i.e.\ they are the $N$ roots of the polynomial of degree $N$ in $z$ that
obtains by multiplying the left-hand side of this equation by $%
\prod\limits_{j=1}^{N} ( z-\zeta _{k} ) $.
\end{lemma}

This lemma  is an immediate consequence of the preceding Lemma~\ref{lemmaA3}, because the secular equation associated with the matrix (\ref{UUU}) (whose roots provide the eigenvalues) is easily seen to coincide (up to an
overall, hence irrelevant, multiplicative constant) with the vanishing of
the determinant in the left-hand side of~(\ref{DetDet}) with (\ref{XnmLemmaA3}) and $x_{k}=\eta _{k}/( z-\zeta _{k})$.

%\textit{Lemma A5}.
\begin{lemma}\label{lemmaA5}
The inverse of the matrix defined componentwise as
follows,
\begin{subequations}
\begin{gather}
M_{nm}=\frac{f_{n} g_{m}}{\xi _{m}-\eta _{n}},\qquad n,m=1,\dots,N,
\label{Mgetaksi}
\end{gather}
is defined componentwise as follows:%
\begin{gather}
\big[ M^{-1}\big] _{nm}  =  \left( \frac{\xi _{n}-\eta _{m}}{g_{n} f_{m}}
\right)  \left[ \prod\limits_{j=1,\; j\neq n}^{N}\left( \frac{\xi _{j}-\eta _{m}
}{\xi _{j}-\xi _{n}}\right) \right]  \left[ \prod\limits_{j=1,\; j\neq
m}^{N}\left( \frac{\eta _{j}-\xi _{n}}{\eta _{j}-\eta _{m}}\right) \right] ,
\notag \\
n,m  = 1,\dots,N.  \label{InverseMnm}
\end{gather}
\end{subequations}
\end{lemma}

The proof of this formula goes as follows. The matrix formula $MM^{-1}=I$,
written componentwise, reads, via~(\ref{Mgetaksi}),
\begin{gather*}
\sum_{k=1}^{N}\left( \frac{g_{k} \big[ M^{-1}\big] _{km}}{\xi _{k}-\eta
_{n}}\right) =\frac{\delta _{nm}}{f_{n}},\qquad n,m=1,\dots,N.
\end{gather*}
Then, for f\/ixed $m$, apply  Lemma~\ref{lemmaA2} with $g_{k}$ replaced by $g_{k}\big[ M^{-1}\big] _{km}$, and $c_{n}$ replaced by $\delta _{nm}/f_{n}$.
This yields, rather immediately, the formula~(\ref{InverseMnm}), which is
thereby proven.

\section{Appendix}\label{appendixB}

In this appendix we detail the derivation of some f\/indings for the fourth
model, f\/irstly the expression (\ref{gnhat4}) of $\hat{g}_{n}(
\underline{z},\underline{\tilde{z}}) $ as solution of the linear
system (\ref{Eqg2}), and secondly the derivation of~(\ref{EqEqMot4}) from~(\ref{EqMotion4bis}).

Via the identity
\begin{gather}
 \frac{1}{ ( \eta  \tilde{z}_{k}+\beta  )   ( \tilde{z}
_{k}-a z_{n}-b ) }=\frac{1}{\eta  ( az_{n}+b ) +\beta }
 \left( \frac{1}{\tilde{z}_{k}-az_{n}-b}-\frac{\eta }{\eta \tilde{z}_{k}+\beta }\right) ,  \label{IdenIden}
\end{gather}
and the def\/inition
\begin{gather}
\sigma ^{( 0) }=\sum_{k=1}^{N}\left( \frac{\eta g_{k}}{\eta \tilde{z}_{k}+\beta }\right) ,  \label{GGG}
\end{gather}
the linear system~(\ref{Eqg2}) can be conveniently reformulated as follows:
%\begin{subequations}
\begin{gather*}
\sum_{k=1}^{N}\left( \frac{g_{k}}{\tilde{z}_{k}-az_{n}-b}\right) =c_{n},
\\
c_{n} =\sigma ^{( 0) }+\frac{\eta ( az_{n}+b )
+\beta }{\eta  z_{n}+\beta }  =\sigma ^{ ( 0 ) }+a+\frac{a}{\eta }\left[ \frac{\beta  (
1-\alpha  ) }{a z_{n}+a \beta  /\eta }\right] .
\end{gather*}
%\end{subequations}

We now use  Lemma \ref{lemmaA2} (with $\xi _{k}=\tilde{z}_{k}$, $\eta
_{n}=a~z_{n}+b$ and $c_{n}$ def\/ined as above). We thus get
%\begin{subequations}
\begin{gather*}
 \hat{g}_{n} ( \underline{z},\underline{\tilde{z}} ) = (
\tilde{z}_{n}-a z_{n}-b )  \left[ \prod\limits_{j=1,\; j\neq n}^{N}\left(
\frac{\tilde{z}_{n}-a z_{j}-b}{\tilde{z}_{n}-\tilde{z}_{j}}\right) \right]
% \notag \\
%\phantom{\hat{g}_{n} ( \underline{z},\underline{\tilde{z}} ) =}{}
%\times
\left[ ( \sigma _{0}+a )  \sigma _{n}^{ ( 1 ) }+
\frac{a\beta ( 1-\alpha ) }{\eta }\sigma _{n}^{( 2)
}\right] ,
\\
\sigma _{n}^{( 1) }=\sum_{k=1}^{N}\left[ \frac{\prod\limits_{j=1,\; j\neq n}^{N} ( \tilde{z}_{j}-az_{k}-b ) }{\prod\limits_{j=1,\; j=k}^{N} ( a z_{j}-a z_{k} ) }\right],\qquad n=1,\dots,N,
\\
 \sigma _{n}^{( 2) }=\sum_{k=1}^{N}\left[ \frac{1}{az_{k}+a\beta /\eta}\frac{\prod\limits_{j=1,\; j\neq n}^{N}(
\tilde{z}_{j}-az_{k}-b)}{\prod\limits_{j=1,\; j=k}^{N} (az_{j}-az_{k}) }\right] ,  \qquad n=1,\dots,N.
\end{gather*}

It is now plain, via the identity (\ref{Identity3a}) (with $\eta
_{k}=az_{k}+b,$ $\zeta _{j}=\tilde{z}_{j})$ that $\sigma _{n}^{(1) }=1$. As for the sum $\sigma _{n}^{( 2) }$, it is also
easily evaluated via the identity (\ref{Identity3}) (now with $\eta
_{k}=az_{k}+b$, $\zeta _{j}=\tilde{z}_{j}$, $z=b-a\beta /\eta )$:
\begin{gather*}
\sigma _{n}^{( 2) }=\frac{a^{-N}\eta }{\eta \tilde{z}_{n}+a\beta -b\eta }\prod\limits_{j=1}^{N}\left( \frac{\eta \tilde{z}
_{j}+a\beta -b\eta }{\eta z_{j}+\beta }\right) .
\end{gather*}
%\end{subequations}
Hence
\begin{subequations}
\begin{gather}
\hat{g}_{n} ( \underline{z},\underline{\tilde{z}} ) = (
\tilde{z}_{n}-az_{n}-b )  \left[ \prod\limits_{j=1,\; j\neq n}^{N}\left(
\frac{az_{j}+b-\tilde{z}_{n}}{\tilde{z}_{j}-\tilde{z}_{n}}\right) \right]
\left( \sigma ^{( 0) }+a+\frac{\sigma ^{( 3) }}{\eta \tilde{z}_{n}+a\beta -b\eta }\right) ,  \label{gnhat}
\\
\sigma ^{( 3) }=a^{1-N}\beta ( 1-\alpha )
\prod\limits_{j=1}^{N}\left( \frac{\eta \tilde{z}_{j}+\alpha \beta }{\eta
z_{j}+\beta }\right) .  \label{sigma3}
\end{gather}
\end{subequations}

Now, using this expression of $\hat{g}_{n}( \underline{z},\underline{\tilde{z}}) $, we can evaluate $\sigma _{0}$ from its def\/inition (\ref{GGG}), thereby obtaining
%\begin{subequations}
\begin{gather*}
\sigma ^{( 0) }=\frac{a\sigma ^{( 4 ) }+\sigma
^{( 3) }\sigma ^{( 5) }}{1-\sigma ^{( 4) }}
=-a+\frac{a+\sigma ^{( 3) }\sigma ^{( 5) }}{1-\sigma
^{( 4) }},  %\label{Eqsigma0}
\\
\sigma ^{(4) }=\sum_{k=1}^{N}\left[ \left( \frac{\tilde{z}
_{k}-az_{k}-b}{\tilde{z}_{k}+\beta /\eta }\right) \prod\limits_{j=1,\; j\neq
k}^{N}\left( \frac{\tilde{z}_{k}-az_{j}-b}{\tilde{z}_{k}-\tilde{z}_{j}}
\right) \right] ,
\\
\sigma ^{( 5) }=\sum_{k=1}^{N}\left[ \frac{\tilde{z}_{k}-az_{k}-b}{( \eta \tilde{z}_{k}+\beta  ) ( \tilde{z}_{k}+\alpha
\beta /\eta  ) }\prod\limits_{j=1,\; j\neq k}^{N}\left( \frac{az_{j}+b-\tilde{z}_{k}}{\tilde{z}_{j}-\tilde{z}_{k}}\right) \right] .
\end{gather*}
%\end{subequations}

It is now plain (via (\ref{Identity3b}) with $\eta _{j}=\tilde{z}_{j},$ $%
\zeta _{j}=a~z_{j}+b,$ $z=-\beta /\eta $) that
\begin{gather}
\sigma ^{( 4) }=1-\prod\limits_{j=1}^{N}\left( \frac{\eta
az_{j}+\eta b+\beta }{\eta \tilde{z}_{j}+\beta }\right) .
\label{sigma4}
\end{gather}

To evaluate $\sigma _{5}$, we use again an identity analogous to that used
above:
\begin{gather*}
 \frac{1}{( \eta \tilde{z}_{k}+\beta ) ( \tilde{z}_{k}+\alpha \beta /\eta ) }=\frac{1}{\beta  ( 1-\alpha )
}  \left( \frac{1}{\tilde{z}_{k}+\alpha \beta /\eta }-\frac{\eta }{\eta \tilde{z}_{k}+\beta }\right).
\end{gather*}
Thereby
\begin{subequations}
\begin{gather}
\sigma ^{( 5) }=\frac{\sigma ^{( 6) }-\sigma ^{(7) }}{\beta ( 1-\alpha ) },  \label{sigma5}
\\
\sigma ^{ ( 6) }=\sum_{k=1}^{N}\left[ \frac{\tilde{z}_{k}-az_{k}-b}{\tilde{z}_{k}+\alpha \beta /\eta }\prod\limits_{j=1,\; j\neq k}^{N}\left(
\frac{az_{j}+b-\tilde{z}_{k}}{\tilde{z}_{j}-\tilde{z}_{k}}\right) \right] ,
\\
\sigma ^{( 7) }=\sum_{k=1}^{N}\left[ \frac{\tilde{z}_{k}-az_{k}-b}{\tilde{z}_{k}+\beta /\eta }\prod\limits_{j=1,\; j\neq k}^{N}\left( \frac{az_{j}+b-\tilde{z}_{k}}{\tilde{z}_{j}-\tilde{z}_{k}}\right) \right] .
\end{gather}
\end{subequations}
Both these sums can be evaluated via the identity (\ref{Identity3b}), with $%
\eta _{k}=\tilde{z}_{k}$, $\zeta _{j}=az_{j}+b,$ and with $z=\alpha \beta
/\eta $ respectively with $z=-\beta /\eta $, obtaining
%\begin{subequations}
\begin{gather*}
\sigma ^{( 6) }=1-a^{N}\prod\limits_{j=1}^{N}\left[ \frac{\eta
z_{j}+\beta }{\eta \tilde{z}_{j}+\alpha \beta }\right] ,  %\label{sigma6}
\qquad
\sigma ^{( 7) }=1-\prod\limits_{j=1}^{N}\left[ \frac{\eta (
az_{j}+b) +\beta }{\eta \tilde{z}_{j}+\beta }\right] ,
%\label{sigma7}
\end{gather*}
%\end{subequations}
hence (via (\ref{sigma5}))
\begin{gather*}
\sigma ^{( 5) }=\frac{1}{\beta ( 1-a) +b\eta }
\left\{ \prod\limits_{j=1}^{N}\left[ \frac{\eta ( az_{j}+b)
+\beta }{\eta \tilde{z}_{j}+\beta }\right] -a^{N}\prod\limits_{j=1}^{N}
\left[ \frac{\eta z_{j}+\beta }{\eta \tilde{z}_{j}+\alpha \beta }\right]
\right\} ,  %\label{sigma5bis}
\end{gather*}
and via this formula together with (\ref{sigma4}) and (\ref{sigma3}) we
f\/inally get
\begin{gather*}
\sigma ^{( 0) }+a=a^{1-N}\prod\limits_{j=1}^{N}\left( \frac{\eta
\tilde{z}_{j}+\alpha \beta }{\eta z_{j}+\beta }\right) .
\end{gather*}

The insertion of this expression of $\sigma ^{( 0) }+a$ and of
the expression (\ref{sigma3}) of $\sigma ^{( 3) }$ in~(\ref{gnhat}) completes the derivation of the expression~(\ref{gnhat4}) of~$\hat{g}_{n}( \underline{z},\underline{\tilde{z}})$.

Finally let us tersely outline the derivation of (\ref{EqEqMot4}) from~(\ref{EqMotion4bis}). Firstly one uses in~(\ref{EqMotion4bis}) the identity (\ref{IdenIden}); then one uses twice the identity~(\ref{Identity3b}), with $%
\zeta _{k}=\widetilde{\tilde{z}}_{k}-b$, $\eta _{k}=a\tilde{z}_{k}$ and with
$z=a  ( az_{n}+b ) $ respectively with $z=-a\beta /\eta $. And
the rest is trivial algebra.

\subsection*{Acknowledgements}

It is a pleasure to thank my colleague and friend Orlando Ragnisco for
pointing out relevant references.

\pdfbookmark[1]{References}{ref}
\LastPageEnding

\end{document}